%% file: draft_v2.tex
\documentclass[traditabstract]{aa} 
\usepackage{txfonts}

\begin{document}
\input psfig.tex

\title{Do X-ray dark or underluminous galaxy clusters exist?}
\titlerunning{Do X-ray dark or underluminous galaxy clusters exist?} 
\author{S. Andreon \and A. Moretti}
\authorrunning{Andreon \& Moretti}
\institute{
INAF--Osservatorio Astronomico di Brera, via Brera 28, 20121, Milano, Italy \\
\email{stefano.andreon, alberto.moretti @brera.inaf.it}
}
\date{Accepted ... Received ...}
\abstract {
We study the X-ray properties of a color-selected sample of clusters at $0.1<z<0.3$, 
to quantify the real aboundance of 
the population of X-ray dark or underluminous clusters and at the same time
the spurious detection contamination level of color-selected cluster catalogs.
Starting from a local sample of color-selected clusters, we restrict our attention to
those with sufficiently deep X-ray observations to probe their
X-ray luminosity down to very faint values and without introducing any
X-ray bias. This allowed us to have an X-ray- unbiased sample of 33 clusters
to measure the L$_X$-richness relation. Swift 1.4 Ms
X-ray observations show that at least 89 \% of 
the color-detected 
clusters are real objects with a potential well deep enough to heat
and retain an intracluster medium. The percentage rises to 94 \% when 
one includes the single spectroscopically confirmed color-selected cluster whose 
X-ray emission is not secured.
Looking at our results from the opposite perspective,
the percentage of X-ray dark
clusters among color-selected clusters is very low: at most about 11 per cent (at 90 \% confidence).
Supplementing our data with those from literature,
we conclude that X-ray- and color- cluster surveys sample the same population and consequently that, 
in this regard 
we can safely use clusters selected with any of the two methods for cosmological purposes. 
This is an essential and promising piece of information  for upcoming surveys in both the 
optical/IR (DES, EUCLID) and X-ray (eRosita).
Richness correlates with X-ray luminosity with a large scatter,
$0.51\pm0.08$ ($0.44\pm0.07$) dex in $lgL_X$ at a given richness,  when $L_x$ is measured
in a 500 (1070) kpc aperture. We release data and software to estimate the X-ray flux, 
or its upper limit, of a source with
over-Poisson background fluctuations (found in this work to be  $\sim 20$ \% on cluster angular
scales)
and to fit X-ray luminosity vs richness if there is an intrinsic scatter.
These Bayesian applications rigorously account for boundaries (e.g. the X-ray luminosity 
and the richness  cannot be negative).
}

\maketitle

\section{Introduction}

Statistical studies of galaxy cluster samples are important for cosmological
studies. The effectiveness of cluster samples
critically depends on several factors, among which is
the accuracy of the knowledge of the
selection function and biases. Cluster samples suited for these
studies are collected by means of surveys in optical, X-ray, or radio
energy bands. Albeit Sunyaev-Zeldovich surveys are now producing samples
useful for cosmological  purposes, currently the most efficient methods to
compile cluster catalogs for cosmological purposes are still based on
optical and X-ray data. Indeed, in the future the two most promising
telescopes in this field seem to be in the X-ray (eRosita) and in the
optical (e.g. DES). 

A critical problem for the use of clusters for cosmological studies
is whether optical and X-ray surveys provide fair samples of the dark
matter halo mass distribution predicted by the perturbation evolution
theories or, equivalently, how far the selection biases in these
surveys are known and under control. An important piece of information
can be provided by the relation between the optical richness ($n200$) and
the X-ray luminosity ($L_X$), which are the fundamental parameters for
cluster detection and are, at the same time, useful mass proxies.
This relation has been previously measured by studying the X-ray properties
of large optically selected cluster samples and it is usually
parametrized by a power law with a (large) intrinsic scatter (Donahue et
al. 2001; Gilbank et al. 2004; Rykoff et al. 2008). The amplitude of the
$L_X$ scatter is commonly explained by the wide range in the dynamical state of
the clusters and by the presence of cooling gas, whereas from the optical
point of view, the scatter can be ascribed to projection effects or
different efficiencies in the galaxy formation (Gilbank et al. 2004). 

The existence of an X-ray dark or underluminous, physically distinct
population, whose X-ray luminosity is much lower than expected from their
optical richness has been also invoked several times in literature. 
The large differences in X-ray and optical
properties of these clusters have been explained by some extreme feedback 
mechanism (e.g. Castellano et al 2011). Recently, Balogh
et al. (2011) studied a  mass-selected sample of 18 moderatly massive (3-6
$10^{14} \ M_{\sun}$) nearby (z $<0.1$) clusters and found a  bimodality in
ICM properties, with a higlhly significant part 
 of X-ray underluminous or dark objects ($\sim$30\%).  
 Therefore, an accurate measurement of
the $L_X-$richness relation and its scatter is surely useful for a better
understanding of the selection biases at different wavelengths and, at the same
time, to probe the cluster non-gravitational physics and the very existence
of X-ray dark or underluminous clusters.

In this work, we study of the X-ray properties and the $L_X$-richness
relation
relation of a small and well-controlled optical  sample. Previous works 
(Donahue et al. 2001; Gilbank et al. 2004; Rykoff
et al. 2008) assembled extensive
cluster optical catalogs and studied their X-ray properties using the
ROSAT shallow observations, mostly the Rosat All Sky Survey.  Their use
of shallow X-ray data (in the cluster rest-frame) resulted in
a large number of almost uninformative upper limits, only ruling out
that the observed cluster has a flux much 
brighter than other similar clusters of the same richness and
which is of little use in ascertaining the existence of 
dark or underluminous clusters. Some of these works (e.g. Rykoff et al.
2008) are also affected by some systematics such as point-source
contamination and centring biases.  
Here we overcome these
limitations through deep X-ray observations of a
well-controlled optically selected cluster sample whose depth is appropriate
to find an X-ray dark population, if this exist.

In Sec. 2 we describe the sample selection; in Secs. 3 and 4 we describe
the optical and X-ray  data analysis. In Section 5 we
describe the fit procedure we used to parametrize the $L_X-$richness
relation. In Section 6 we revisit previous statements about the
existence of underluminous clusters. In
sec. 7 we briefly discuss our results. We summarize and 
conclude our work in Section 8.
Throughout this paper we assume $\Omega_M=0.3$, $\Omega_\Lambda=0.7$  and
$H_0=70$ km s$^{-1}$ Mpc$^{-1}$. For the statistical analysis, we adopt a 
Bayesian framework with uniform priors, unless otherwise stated.

\section{Sample selection}

We start from the maxBCG cluster catalog (Koester et al. 2007), which is an optically selected,
quasi-volume-limited sample of clusters with $0.1<z<0.3$, with very accurate photometric redshifts
($\delta z \sim 0.01$). 

We searched for all Swift X-Ray Telescope (XRT hereafter) observations within 8 arcmin from any
maxBCG cluster and with an exposure time longer than 3 ks. This yielded 180 observations
out of 14000  in the Swift archive at the start of this work.

We restricted our analysis to the clusters with high-quality X-ray observations. 
We keep those, of the 180 selected clusters, whose
3$\sigma$ flux limit is at least 30 times fainter than the expected X-ray cluster flux.
We calculated the $3\sigma$ flux limit (for a point-like source) as in Moretti et al. (2007) 
and the expected X-ray flux 
assuming the $L_X-n200$ relation reported in Rykoff et al. (2008)\footnote{ $n200$ is a measure of the
cluster richness, and it is given by the
number of cluster galaxies measured in some standard conditions,
see Sec 3 for details.}.
This filter can be expressed by
\begin{equation}
\log f_{lim}< 42.22 + 1.82 \log \frac{n200}{40} -d_L^2,  
\end{equation}
where the flux limit is expressed in erg s$^{-1}$ cm$^{-2}$ and $d_L$ is the cluster  luminosity distance in cm.
This selection is such that an {\it average} cluster of a given richness $n200$ is 
very well detectable in X-ray and allows us to 
exclude most uninformative upper limits. 
For example, the poorest cluster cataloged in maxBCG catalog (richness $n200=10$),
at the typical redshift of our sample, $z=0.2$, 
is included in the sample only if it has been observed for at least 15 ks with the 
Swift XRT, corresponding to a $3\sigma$ flux limit
of 2 10$^{-14}$erg s$^{-1}$ cm$^{-2}$.
Equation 1 leaves us with 43 clusters. We emphasize that a cluster was kept or removed 
from the sample
independently of its own $L_X$ (in this phase, X-ray data were not even downloaded), which is
essential to avoid X-ray biases, as discussed in Sec. 6.

In the XRT archive we found three clusters, MS1006+1202, MS1455.0+232 and Abell 1835,
which are the target of the observations.
To search for underluminous clusters they are not useful, because they
are known to be X-ray-bright sources, and leaving them in the sample would introduce a bias in the
$L_{X}-$richness relation.
Therefore we removed them from the statistical sample, but we kept them in tables and figures. 

Three more clusters 
are aligned with an unrelated bright point X-ray sources, making the measurement of the cluster X-ray emission useless for our
purposes. We discarded them. Two more clusters 
fall too near to the Swift XRT field-of-view boundary to make the X-ray data reliable. 
We discarded these as well. Finally,
two maxBCG clusters are (or might be) multiple detections of clusters already present 
in the maxBCG catalog. 
To avoid any ambiguity, we also discarded these, which left us with a final sample of 33 (+3) 
clusters.

Again we stress that these selections do not introduce any selection effect on the X-ray axis:
the cluster X-ray flux is not used, directly or indirectly, to decide if the cluster has to be
kept in the sample. The final cluster sample is formed by either serendipitously observed 
clusters (in the field of
a source at a fairly different redshift), or, in 50 \% of the cases, belong to our own Swift
observational program
targeting all rich maxBCG clusters. 

\input table2.tex

\section{Cluster center, richness, mass}

We started from the maxBCG catalog to improve the center and richness of our clusters.

The maxBCG catalog reports the coordinates of the brightest galaxy (BCG) in
the region as the cluster center. In 10 cases 
(\# 3, 4, 7, 12, 19, 21, 23, 34, 37, 38), the BCG has been mis-identified
in the maxBCG catalog, and the quoted cluster center is 
offset by both the peak of the galaxy density and  by the X-ray 
emission barycenter (which is centered on the galaxy overdensity, see Sec 4) 
by more than 30 arcsec. Therefore, we updated the cluster center.
We note that the fraction of miscentered clusters, $0.28\pm0.07$, 
derived from our observations of 10 offset in a sample of 33, agrees with 
the rough expectations based on simulations (Johnston et al. 2007; Hilbert \& 
White 2010). We emphasize that both the value and the error of this fraction are 
essential parameters for estimating
cosmology parameters (Hilbert \& White 2010) or for
forecasting their precision in future surveys (Oguri \& Takada 2010) using
galaxy clusters. Up to now, a rough estimate for the value was taken, and no error on it
was considered. The values directly measured for the first time here on real data 
allow future analyses to provide more realistic estimates.

We derived the cluster richness, $n200$, using the Sloan Digital Sky Survey (hereafter SDSS) $6^{th}$ data 
release (Adelman-McCarthy et al., 2008), strictly following the Andreon \& Hurn (2010) procedure, which 
rigorously accounts for the finite sample size, uncertainties, and existence of boundaries (e.g. clusters galaxies
do not come in negative units, while the usual total minus background difference may be negative because of 
Poisson fluctuations). 
We counted the net number of red galaxies within $r_{200}$. This radius is estimated from the net number 
of red galaxies within $1.43$ Mpc from the cluster center, $obsn(<1.43)$, using equation 18 in Andreon \& Hurn (2010),
which calibrates this relation with a sample of 54 clusters with kwown $r_{200}$. 

Moreover, we used the Andreon \& Hurn (2010) measured richness--mass scaling to estimate the masses of our clusters.
The quoted mass uncertainty accounts for a number of error sources including the larger calibration uncertainty 
at the extremes of the richness range, as detailed in Andreon \& Hurn (2010). This point has relevance
for the richest cluster of our sample, \# 38, which has a larger mass error because in the calibrating sample only 
few clusters are as rich as it is. At the other richness extreme, the error of the poorest cluster in our sample, 
\#9, accounts for the extrapolation in going from the range where the richness-mass is well calibrated,
from seven galaxies on, to its richness, about four galaxies. The model is described in detail in Andreon \& Hurn (2010), 
who also give its coding in a user-friendly way.

Table 1 lists the results of the optical analysis. Column 1 lists the 
cluster id; column 2 and 3 list updated coordinates; 
column 4 lists the net observed number of
galaxies in the cluster line-of-sight within an aperture of 1.43 
Mpc, $obsn(<1.43)$; column 5 lists the observed number of
galaxies in the cluster line of sight within $r_{200}$, $obsgaltot_i$;
column 6 gives the
observed number of galaxies in the background line-of-sight 
$obsgalbkg_i$; column 7 lists the
ratio between the cluster and background solid angles, $Cgal$;
column 8 gives
the cluster richness (posterior mean and highest posterior 68 \% interval); 
column 9 gives
the inferred mass (posterior mean and standard deviation), on a log
scale in solar mass units. Finally, column 10 lists other
known identifications of the studied clusters when their reported
coordinates is within 1.5 arcmin from the center determined by us.
The angular offset is reported in parenthesis.

Figure 1 shows the distribution in mass of clusters in our sample: most of them
are in the range 1 to 5 $10^{14}$ solar masses, and all are included in
the range $0.6$ to $8$ $10^{14}$ solar masses. 

\begin{figure}
\centerline{\psfig{figure=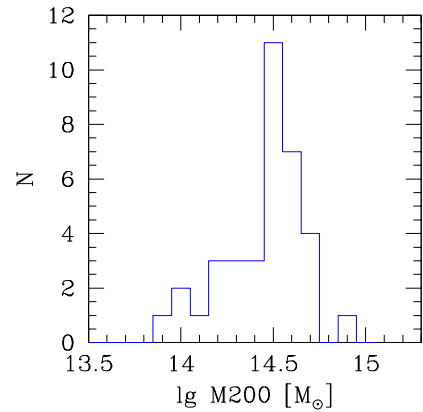,width=6truecm,clip=}}
\caption[h]{Distribution of cluster masses.
}
\end{figure}

\begin{figure}
\centerline{\psfig{figure=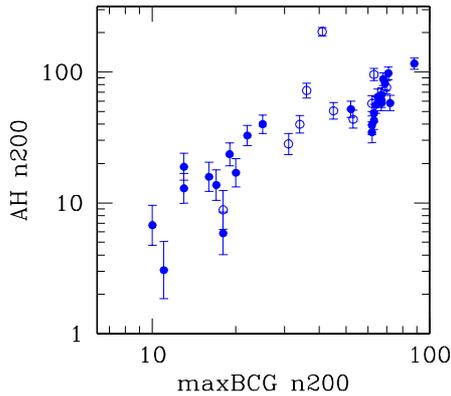,width=6truecm,clip=}}
\caption[h]{Comparison of the richness as determined by us
(ordinate) and in the maxBCG catalog. Open points
marks miscentered clusters in the maxBCG catalog. 
There is
no error on the abscissa because none is listed in the maxBCG catalog.
}
\end{figure}

Figure 2 compares the Koester et al. (2007) and our measurements of 
richness.
There is no errorbar in Figure 2 on maxBCG richness because none
is listed in their catalog. 
Figure 2 shows that our sample explores a richness range that goes from  
the richest (maxBCG richness $\sim 80$) to the poorest
(maxBCG richness $10$) clusters in the $0.1<z<0.3$
volume (and in the SDSS area).
The two richness estimates broadly agree, although they were
derived in slightly different ways: a)
Koester et al. count galaxies in different
color and luminosity ranges and in different filters; b)
we account for background galaxies, whereas Koester et al.
do not; c) we adopt 
different centers (in 30 \% of the cases)
and also $r_{200}$ values (Koester et al. count
galaxies within a radius, unfortunately named $r_{200}$, 
which is on average $2 r_{200}$,
e.g. Sheldon et al. 2009; Becker et al. 2007; Johnston et al. 2007). 
Therefore, it is not
surprising that a scatter is present between the two richnesses.
The outlier point in Fig. 2 (the point with highest ordinate) is
cluster \#38, and it is one of those that have
a wrong maxBCG estimate
of the cluster center. In particular, the area explored by maxBCG
(i.e. the circle centered on their center and of radius given by
their $r_{200}$) misses about half the cluster.

\begin{figure}
\centerline{\psfig{figure=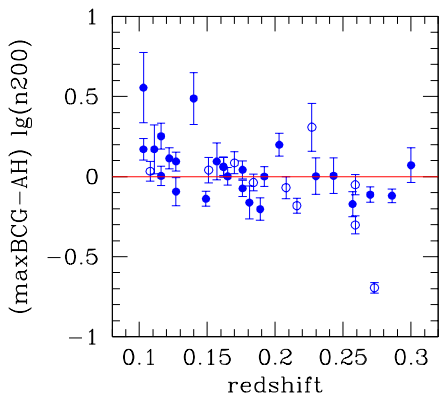,width=6truecm,clip=}}
\caption[h]{Redshift dependence of richness residuals. Open points
indicate miscentered clusters in the maxBCG
catalog. Error bars
only consider errors on our richness estimate, because there is no such
measurement for the maxBCG.
}
\end{figure}

Figure 3 compares residuals between Koester et al. (2007, maxBCG) and our (AH) 
richnesses vs redshift. There is a small but clear trend with redshift, in the
sense that
maxBCG richnesses are underestimated at high redshift. Indeed,
residuals tend to be positive for
the lower half of the redshift range and negative for
the upper half.
The redshift dependency of the maxBCG richness has already
been indirectly pointed out
(Reyes et al. 2008, Rykoff et al. 2008;
Becker et al. 2007;
Rozo et al. 2009).
By directly comparing two richness estimates, Figure 3 confirms
that the maxBCG richness is redshift-dependent.
A redshift trend introduces a systematic
bias on the mass estimate and therefore 
on estimates of cosmological parameters.

\begin{figure*}
\centerline{%
\psfig{figure=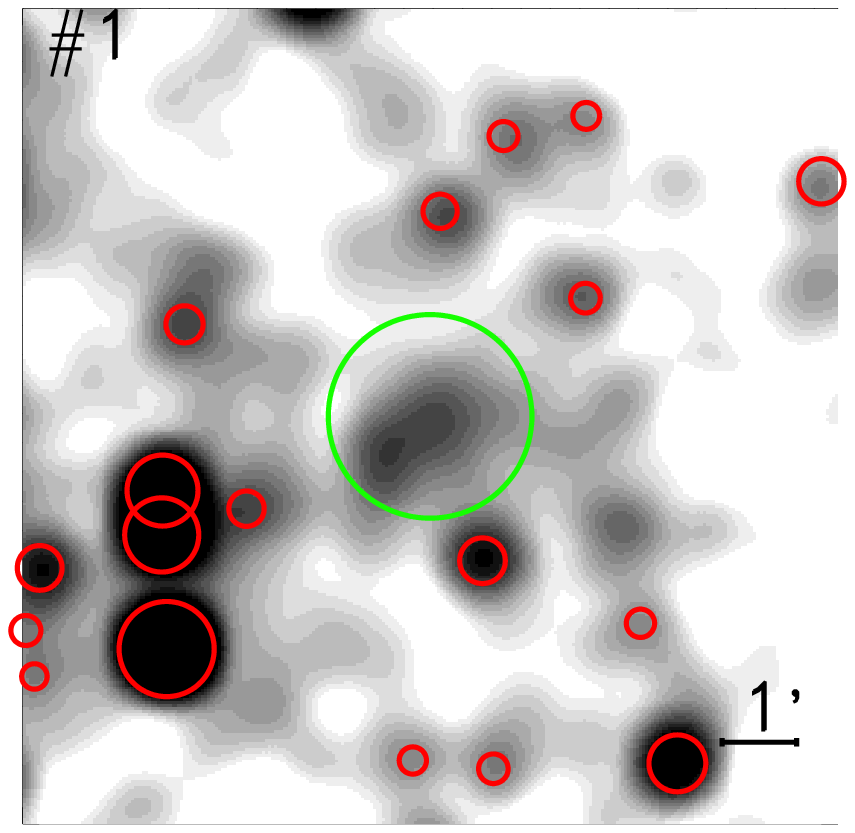,width=2.8truecm,clip=}
\psfig{figure=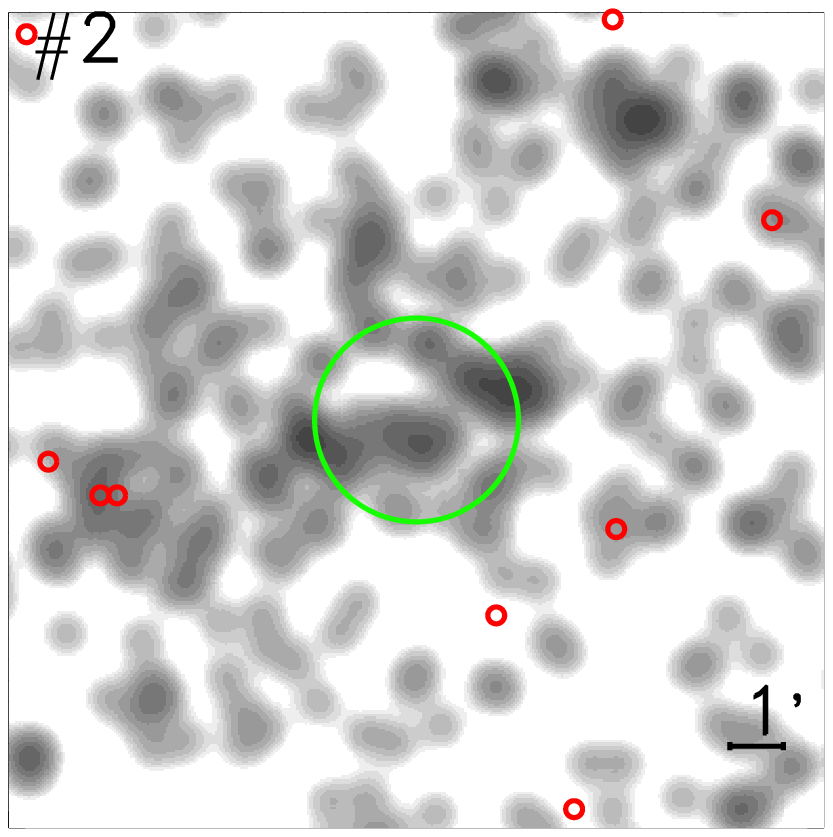,width=2.8truecm,clip=}
\psfig{figure=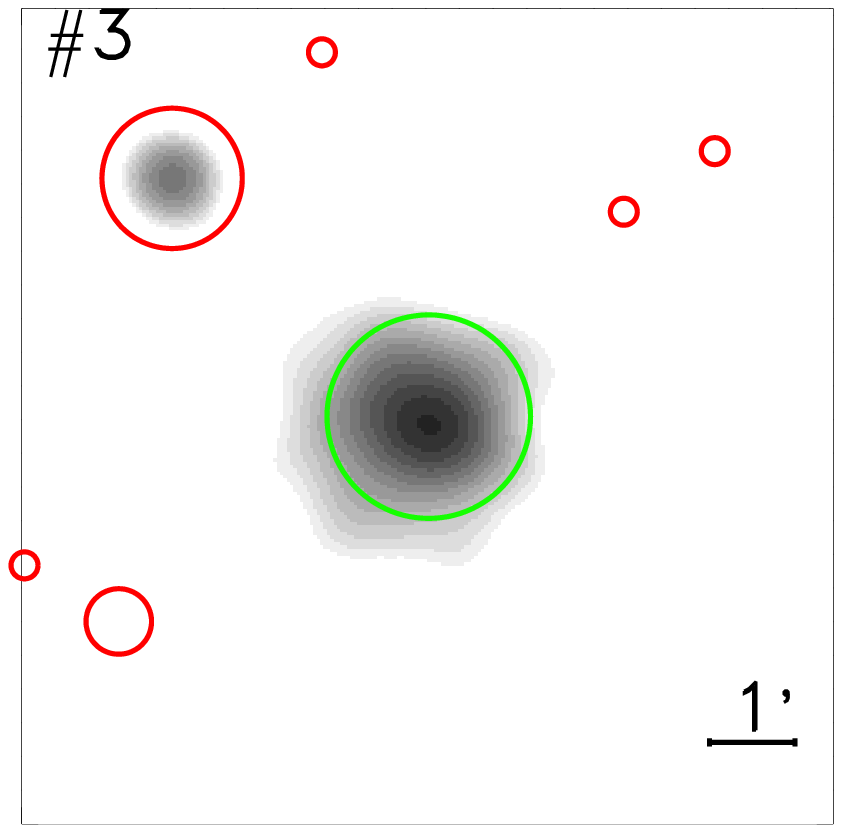,width=2.8truecm,clip=}
\psfig{figure=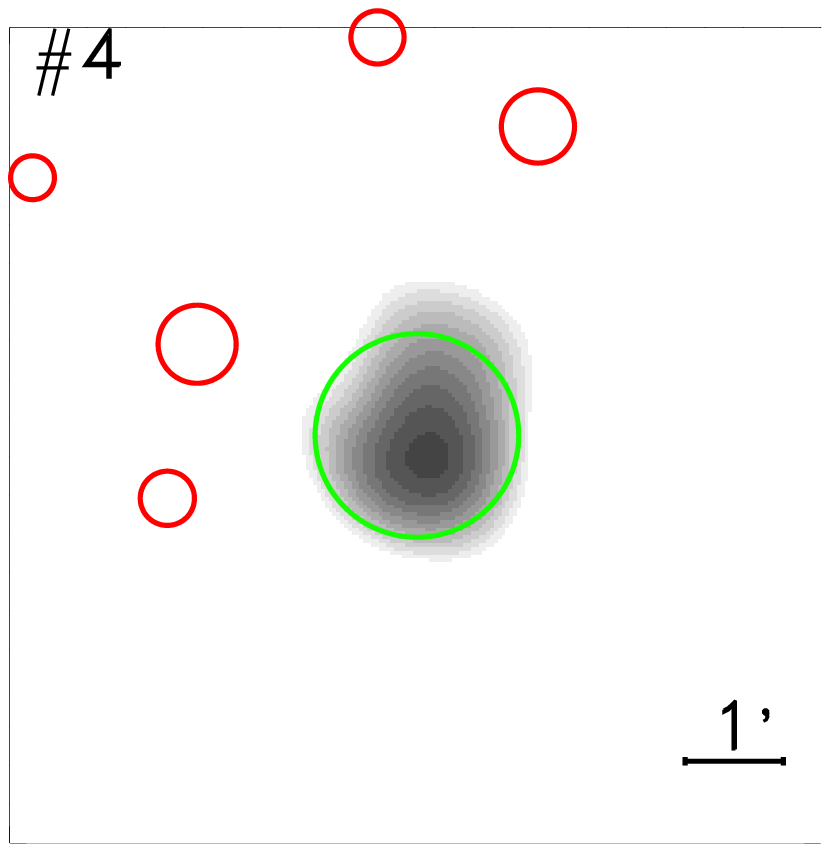,width=2.8truecm,clip=}
\psfig{figure=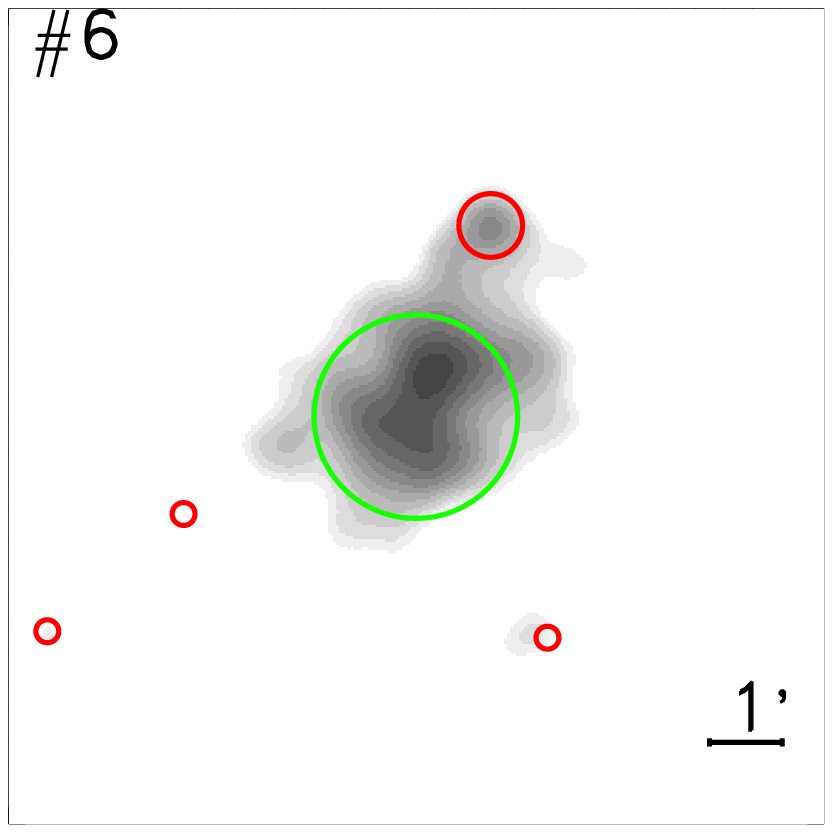,width=2.8truecm,clip=}
\psfig{figure=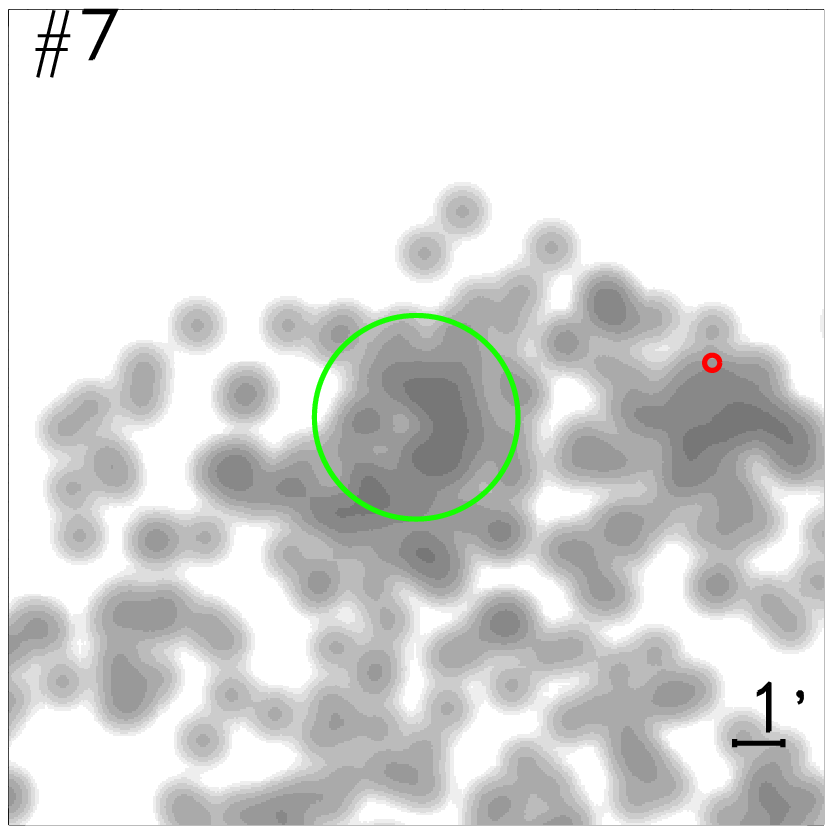,width=2.8truecm,clip=}
}
\centerline{%
\psfig{figure=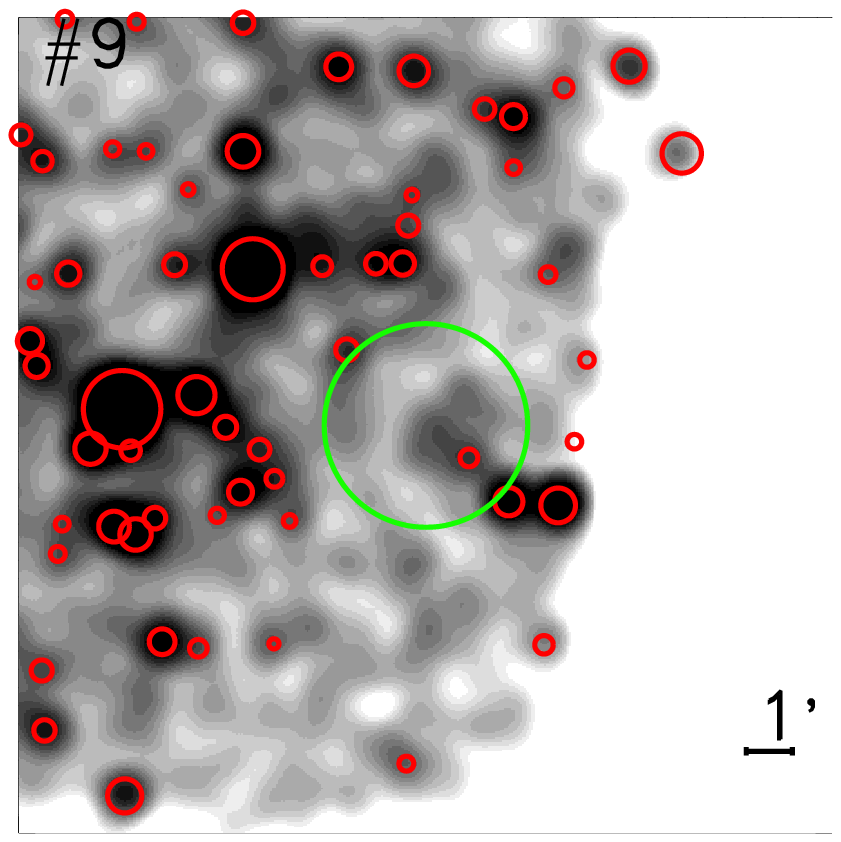,width=2.8truecm,clip=}
\psfig{figure=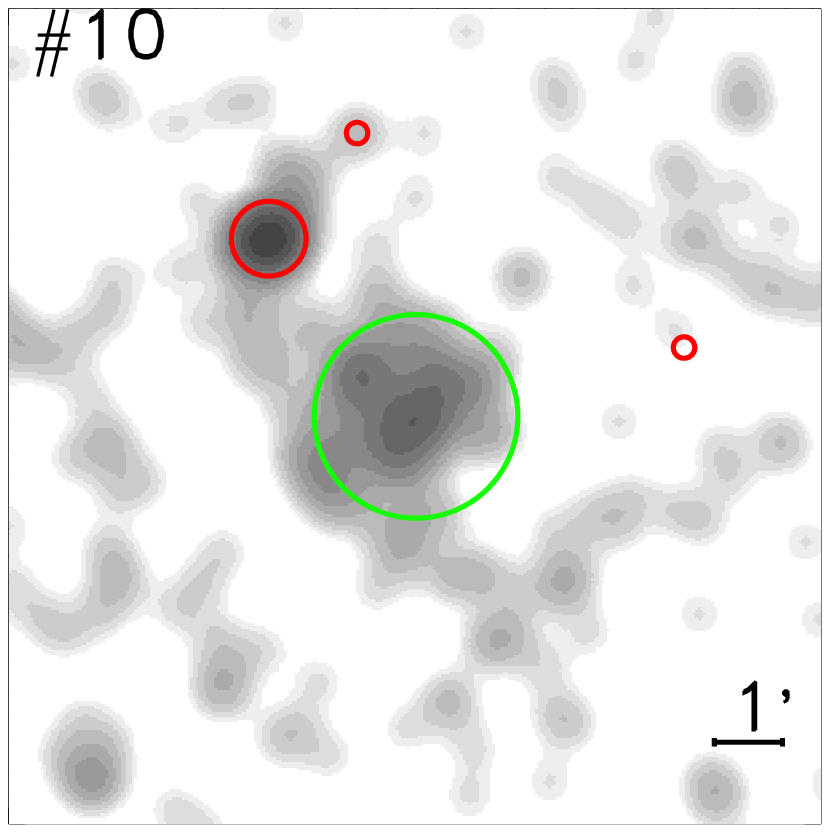,width=2.8truecm,clip=}
\psfig{figure=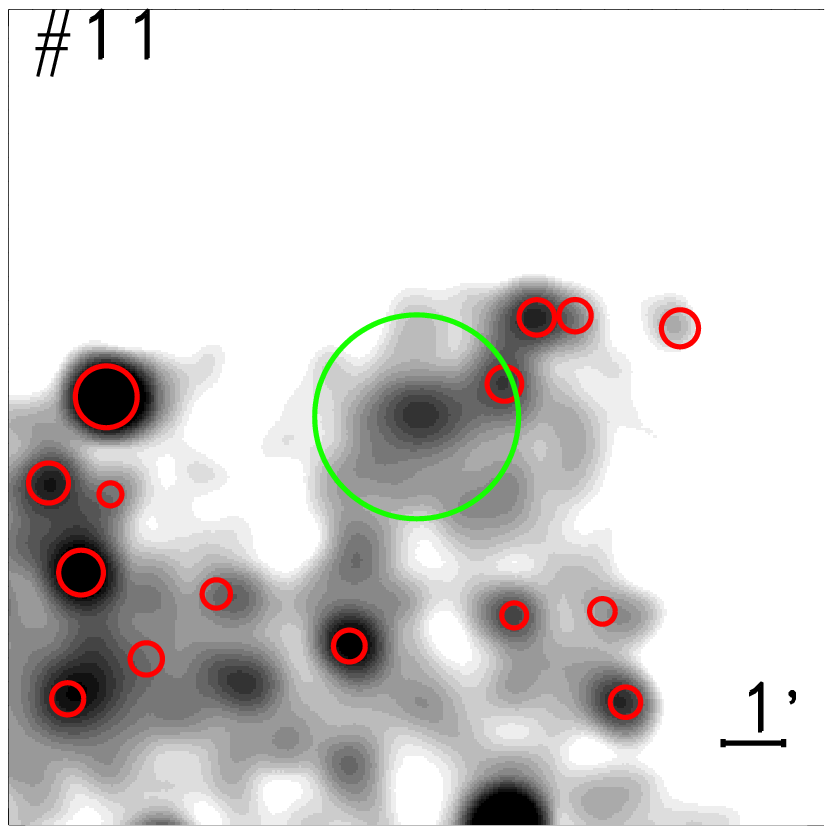,width=2.8truecm,clip=}
\psfig{figure=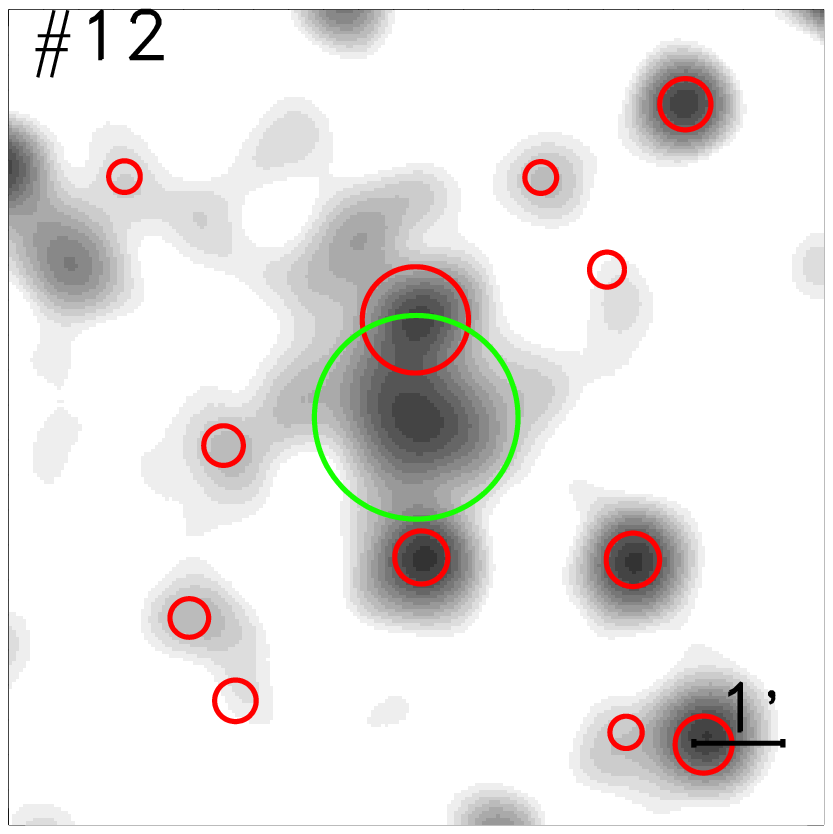,width=2.8truecm,clip=}
\psfig{figure=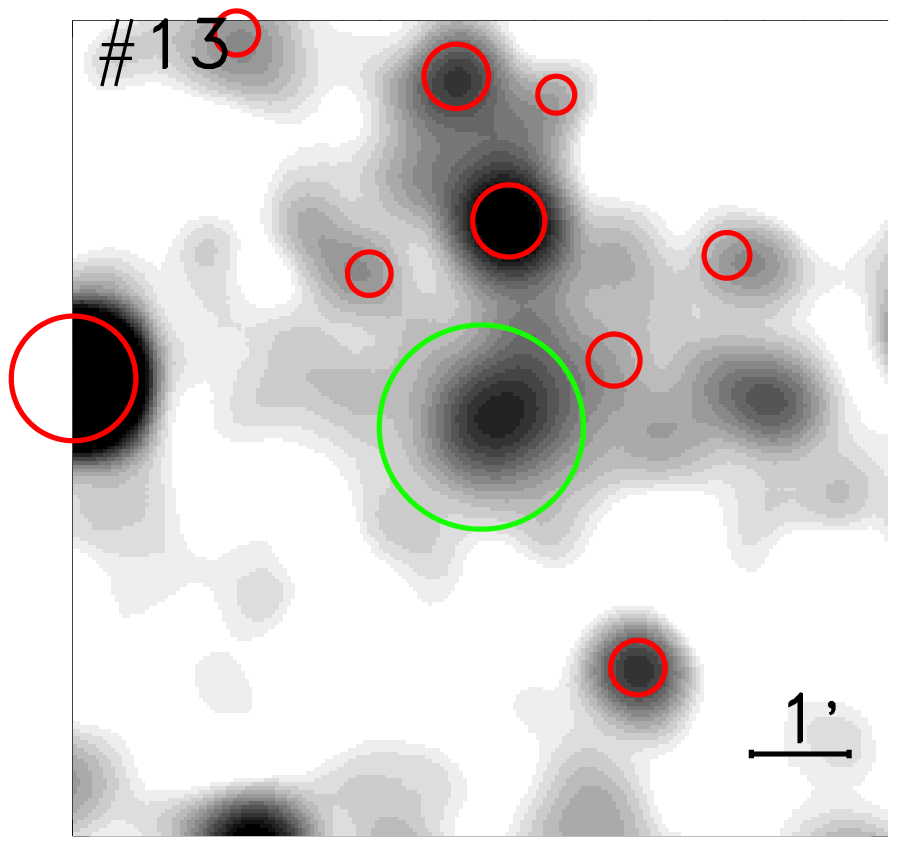,width=2.8truecm,clip=}
\psfig{figure=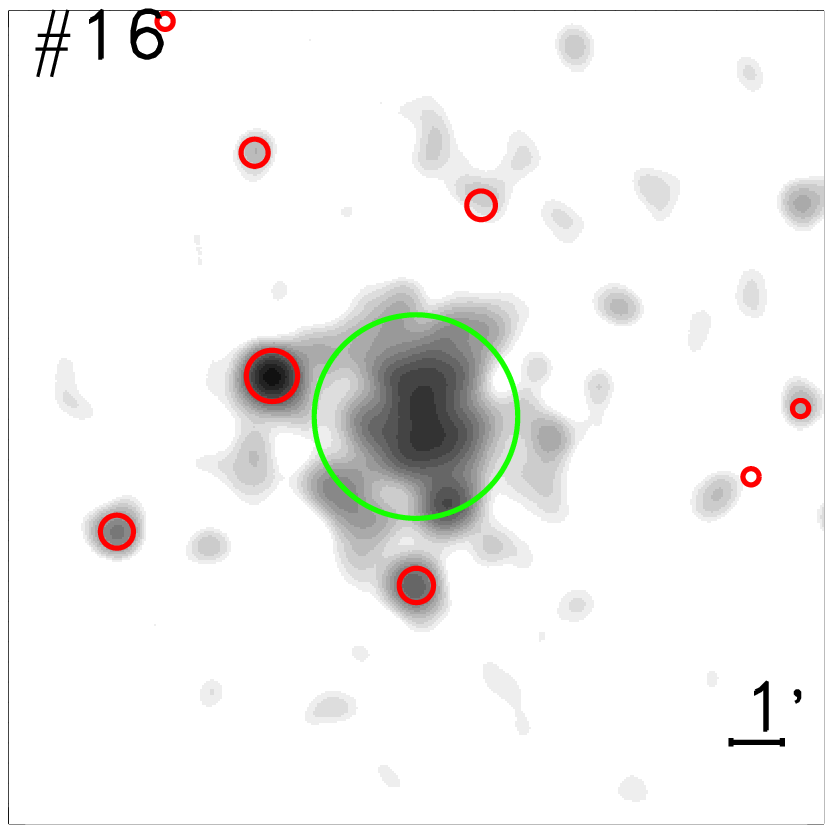,width=2.8truecm,clip=}
}
\centerline{%
\psfig{figure=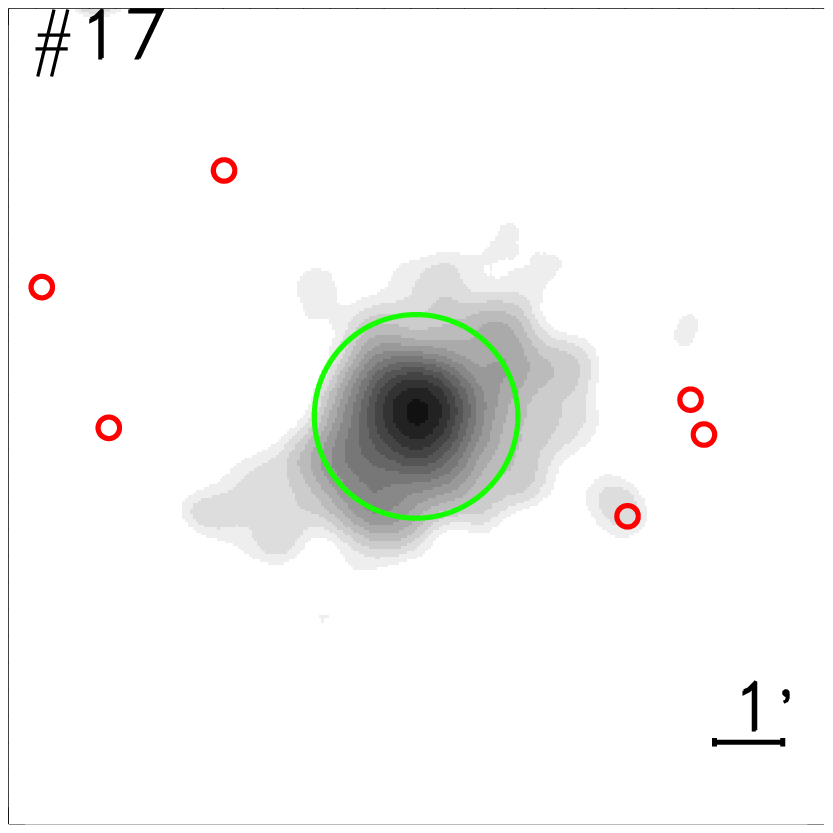,width=2.8truecm,clip=}
\psfig{figure=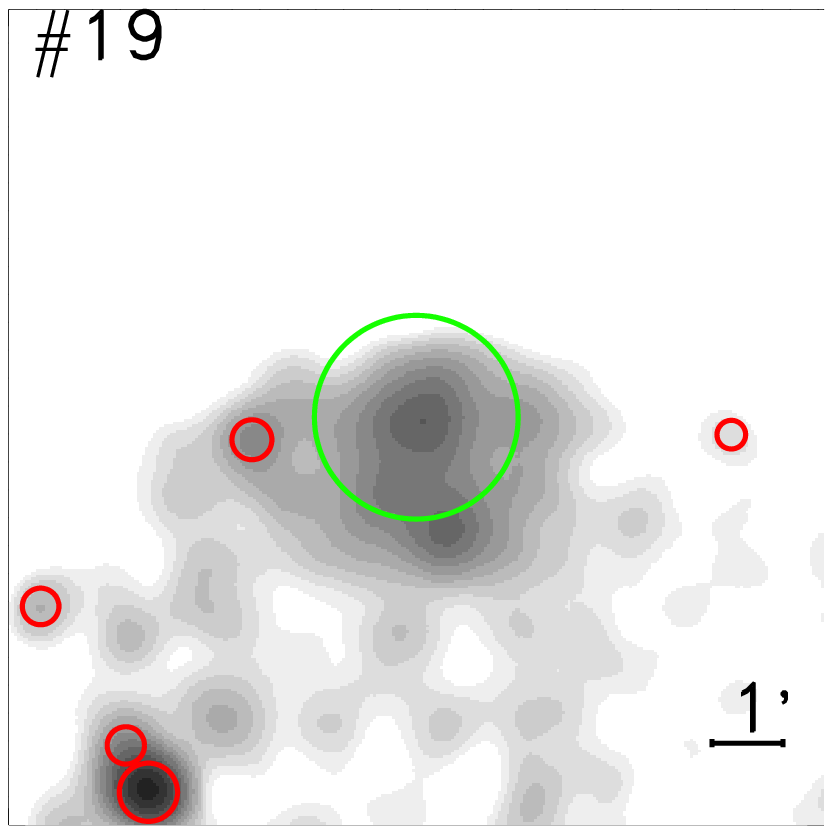,width=2.8truecm,clip=}
\psfig{figure=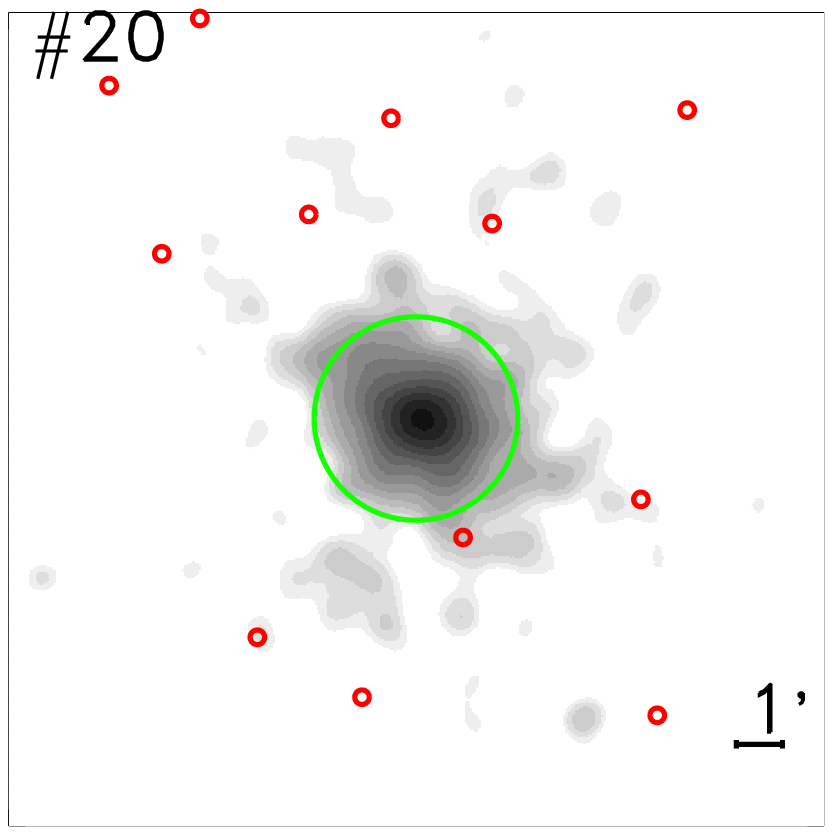,width=2.8truecm,clip=}
\psfig{figure=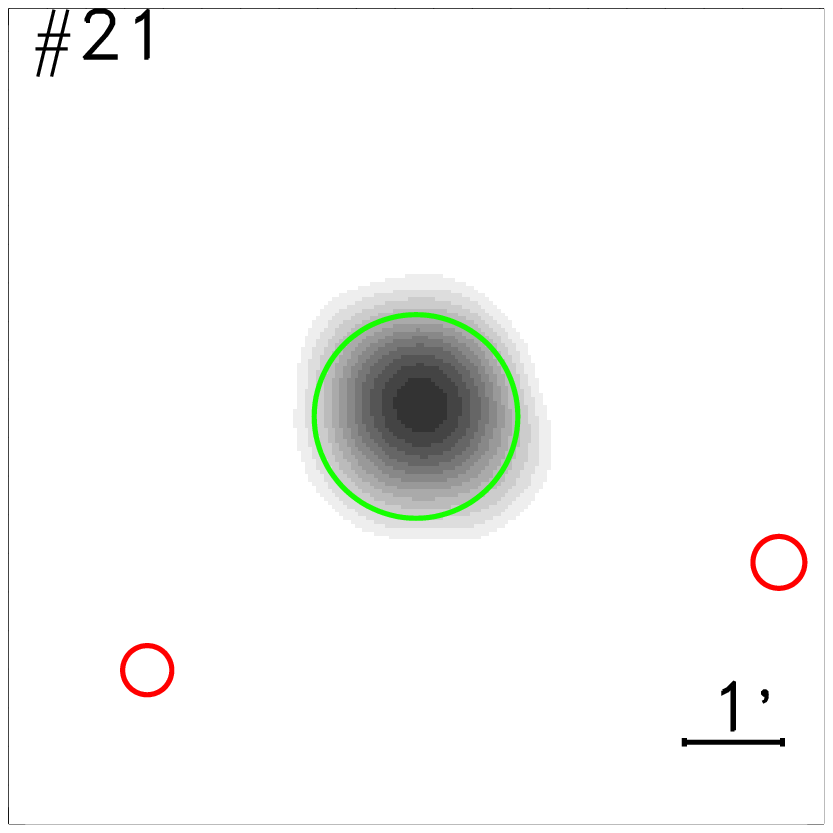,width=2.8truecm,clip=}
\psfig{figure=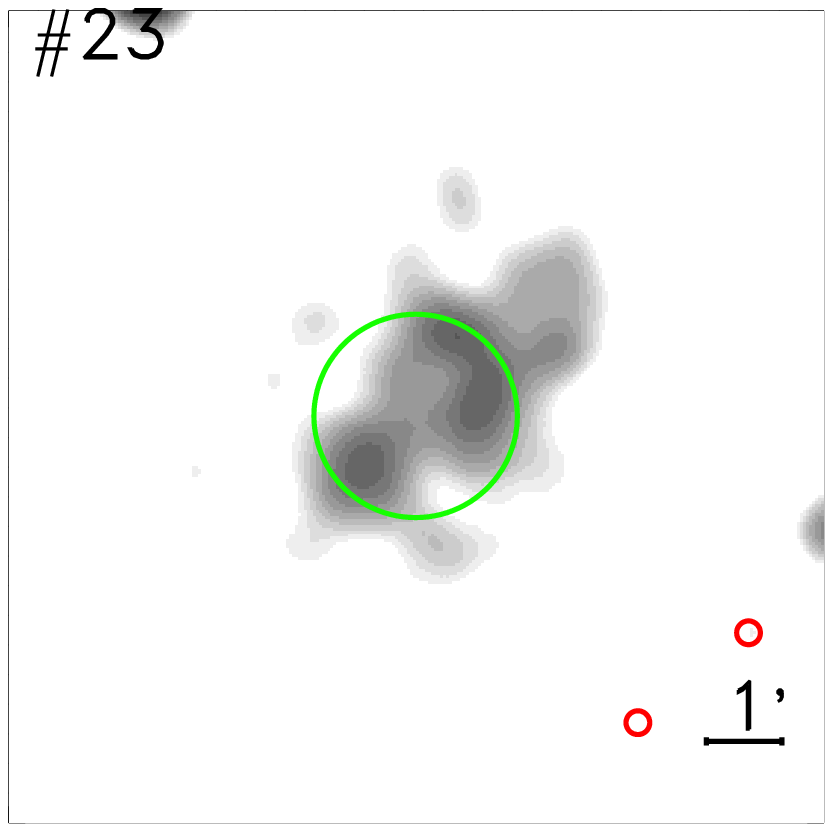,width=2.8truecm,clip=}
\psfig{figure=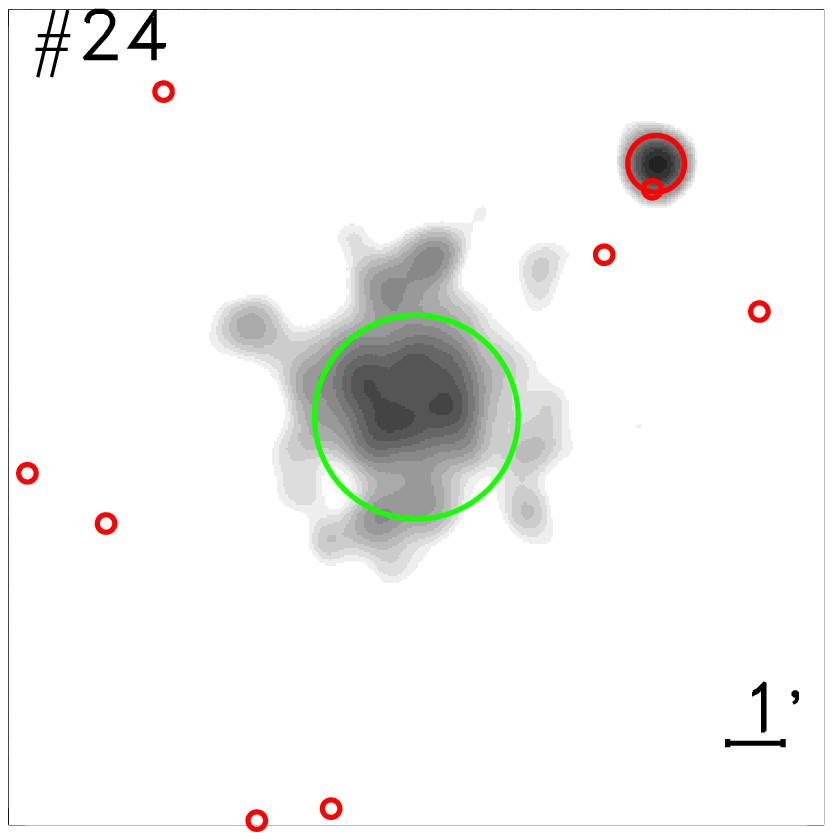,width=2.8truecm,clip=}
}
\centerline{%
\psfig{figure=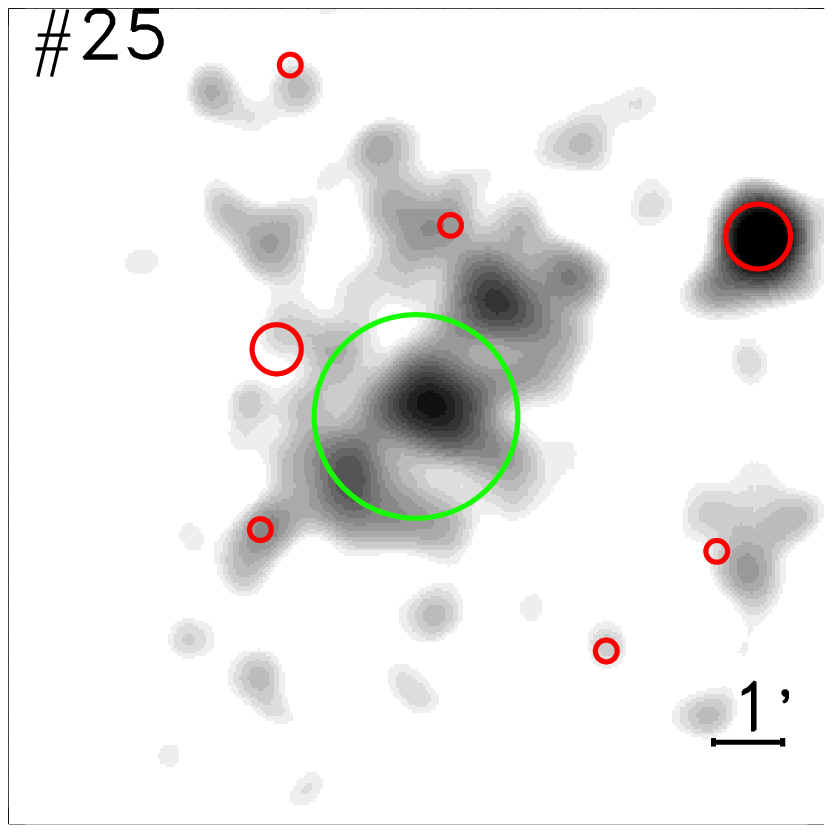,width=2.8truecm,clip=}
\psfig{figure=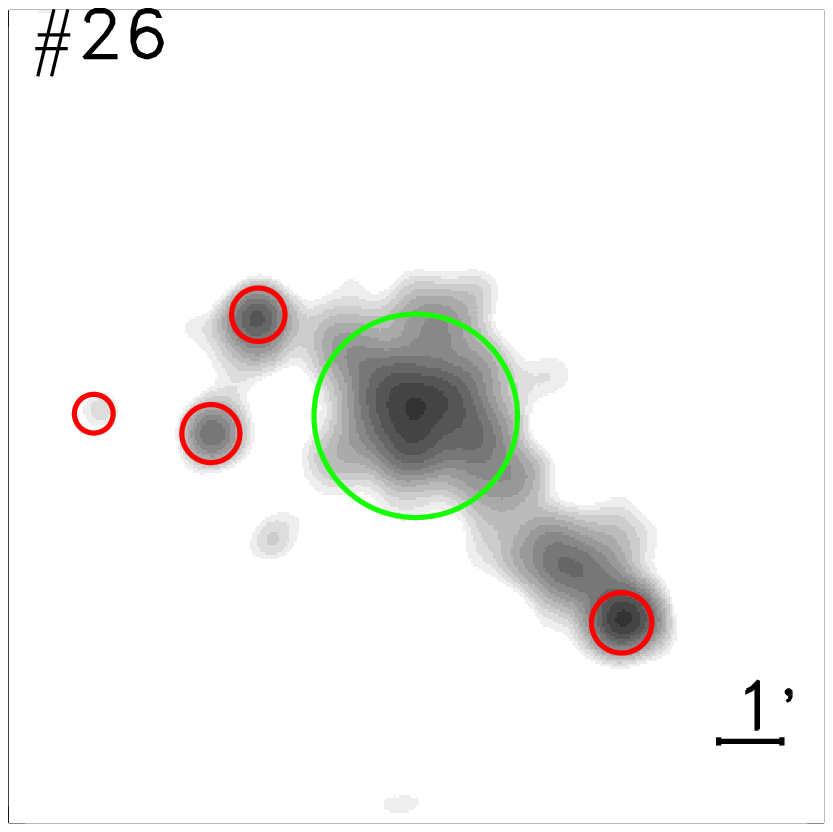,width=2.8truecm,clip=}
\psfig{figure=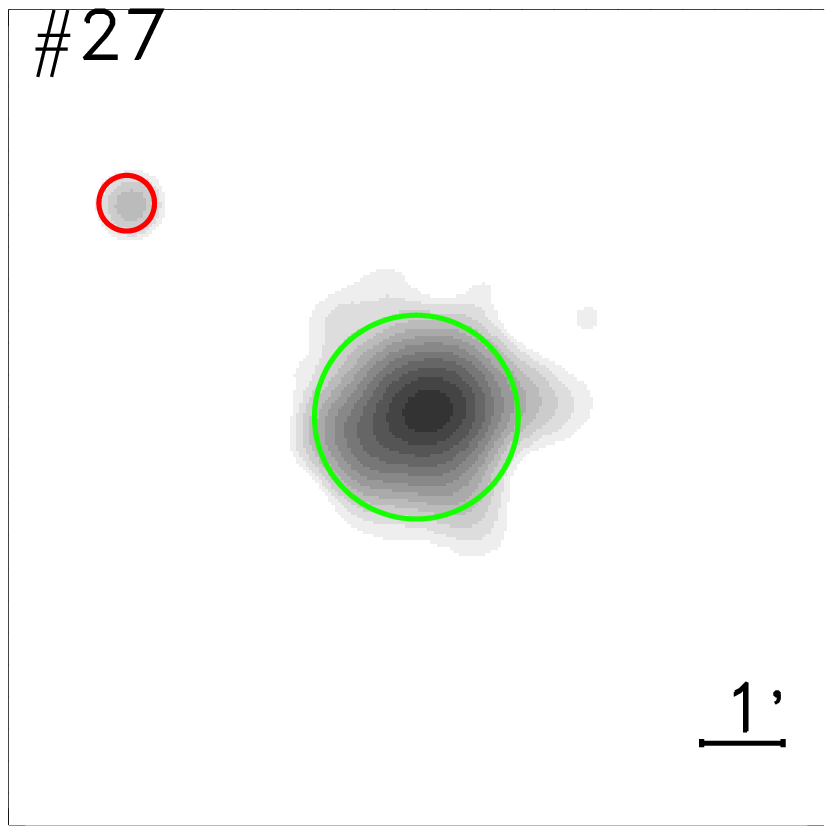,width=2.8truecm,clip=}
\psfig{figure=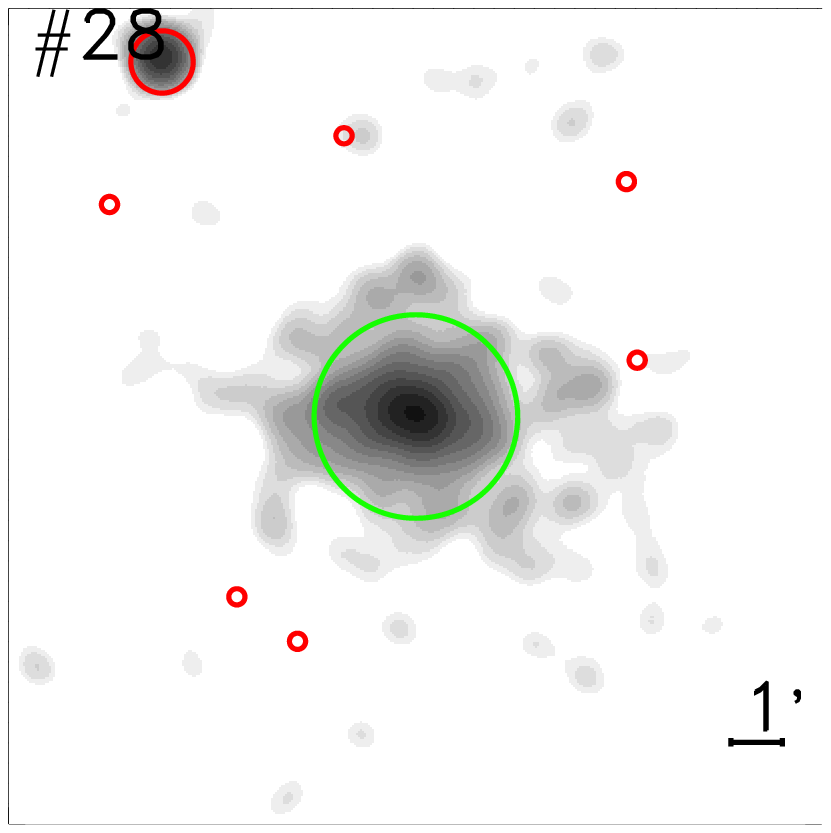,width=2.8truecm,clip=}
\psfig{figure=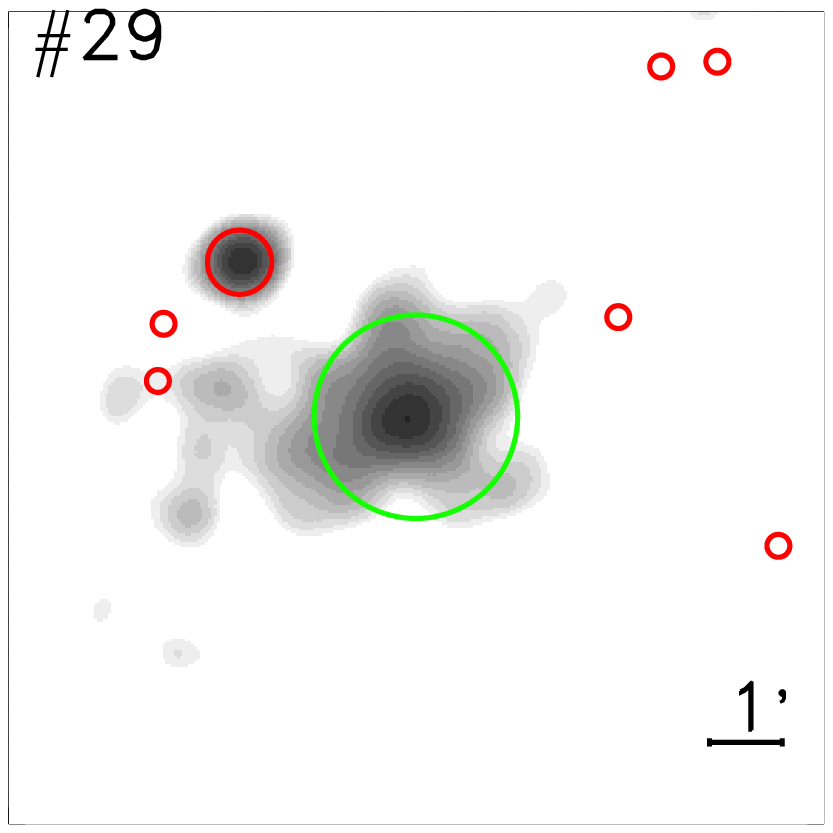,width=2.8truecm,clip=}
\psfig{figure=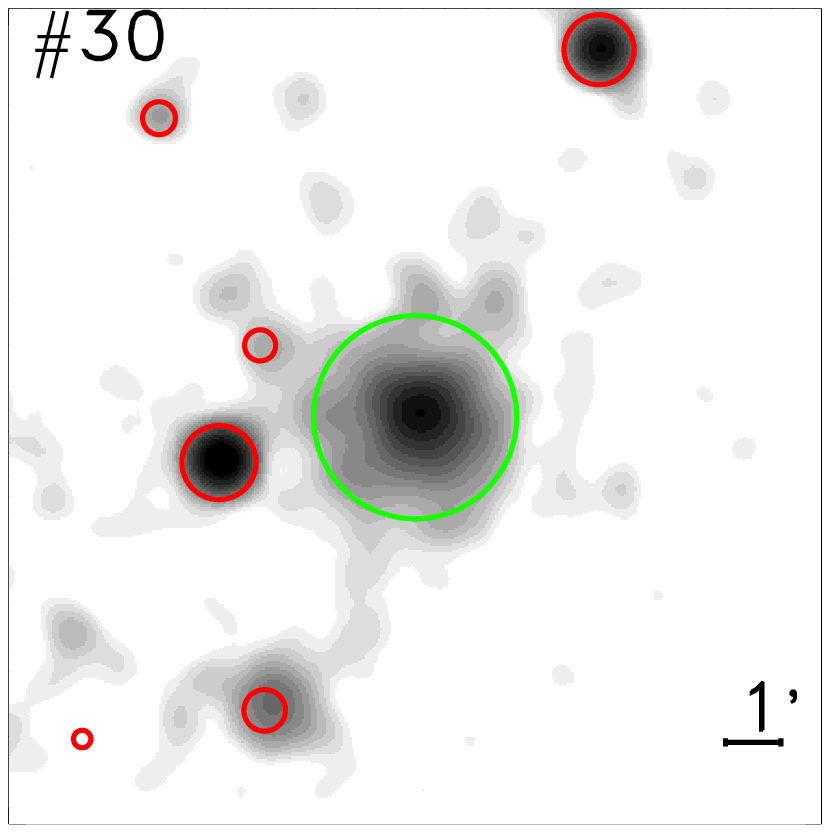,width=2.8truecm,clip=}
}
\centerline{%
\psfig{figure=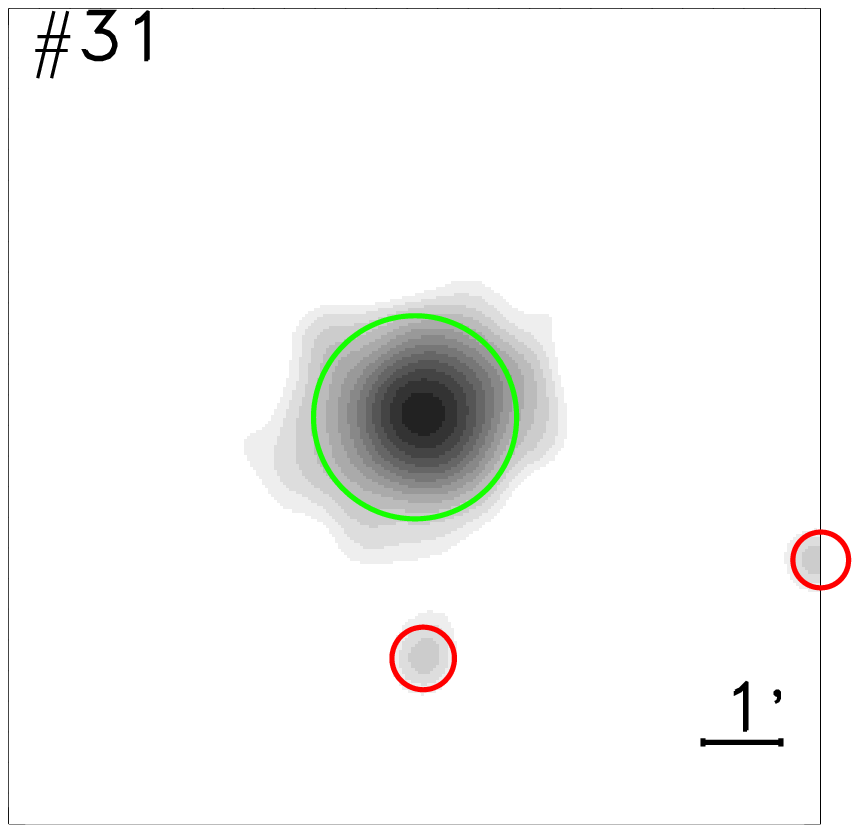,width=2.8truecm,clip=}
\psfig{figure=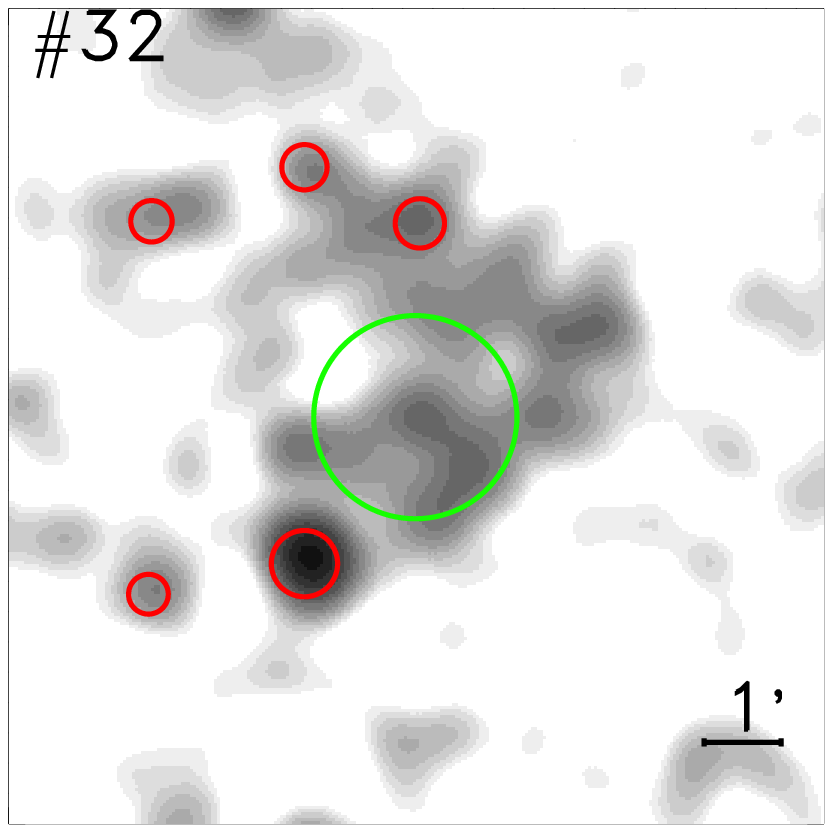,width=2.8truecm,clip=}
\psfig{figure=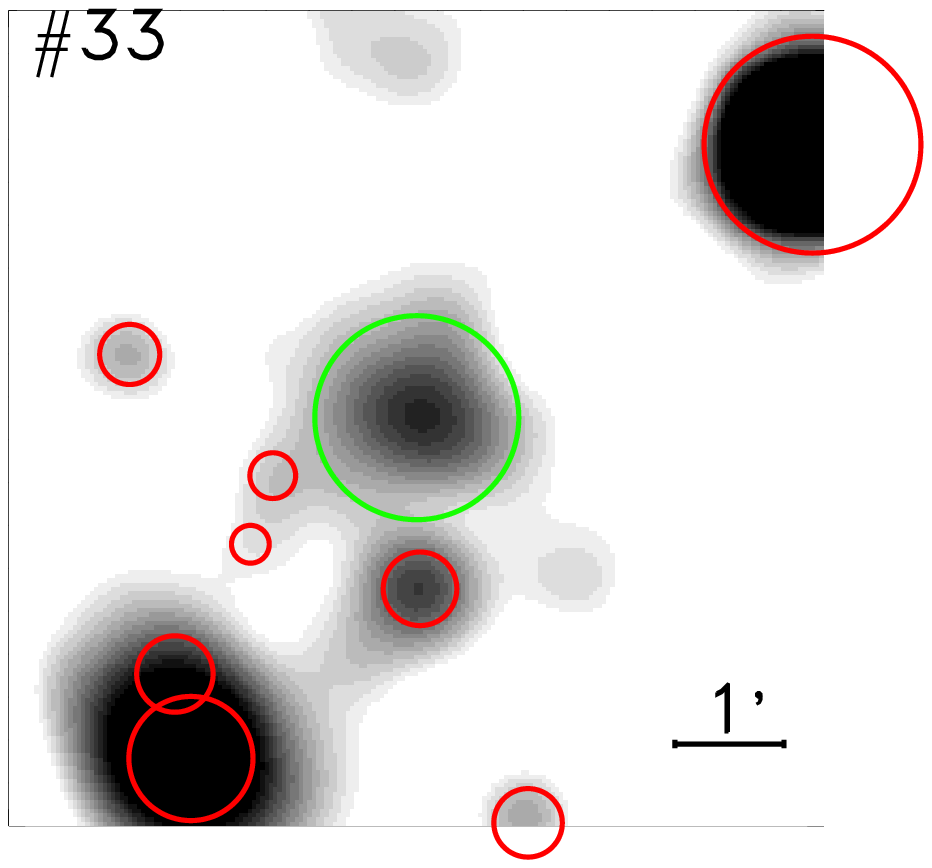,width=2.8truecm,clip=}
\psfig{figure=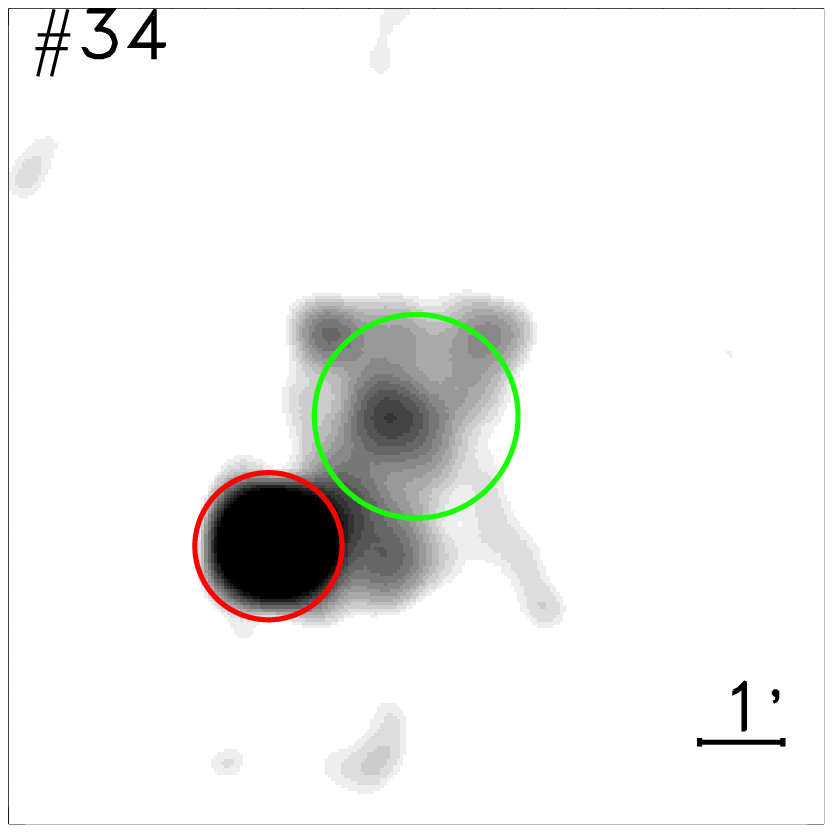,width=2.8truecm,clip=}
\psfig{figure=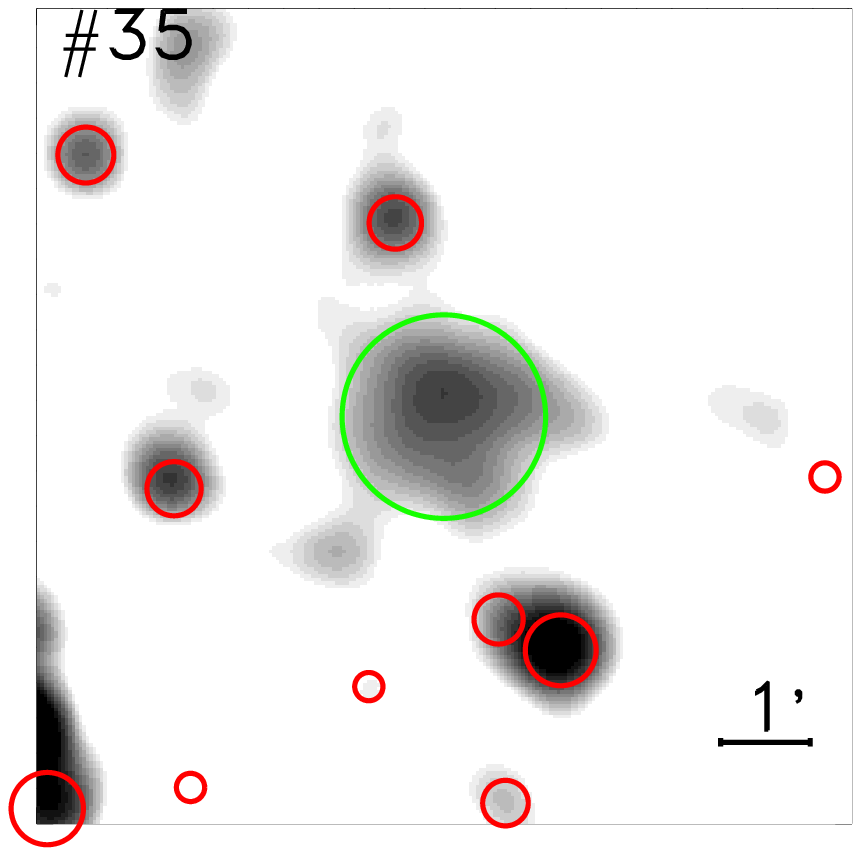,width=2.8truecm,clip=}
\psfig{figure=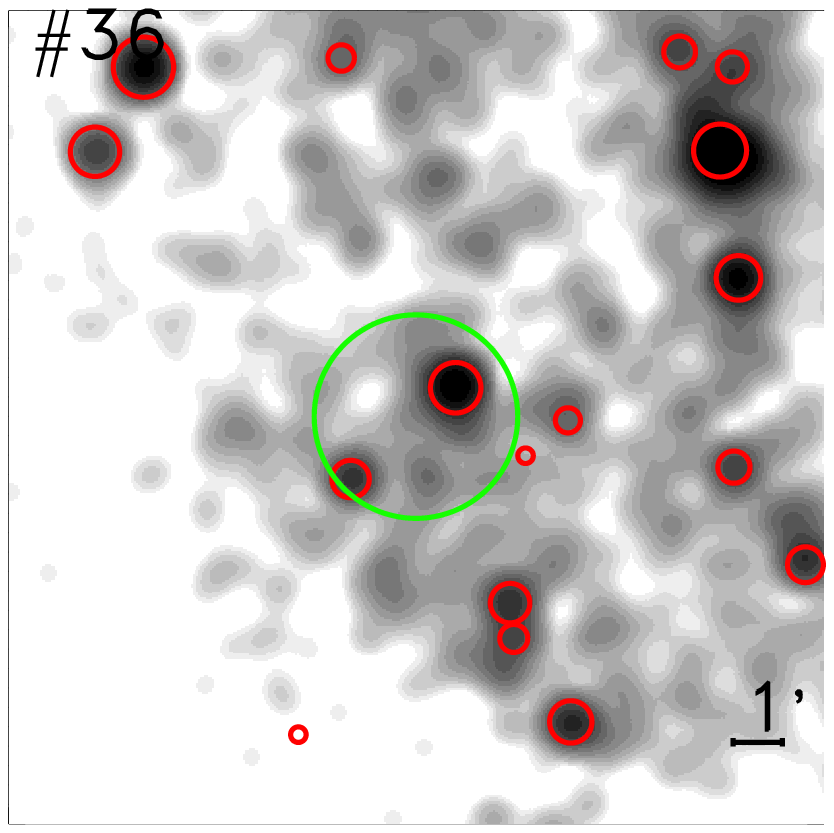,width=2.8truecm,clip=}
}
\centerline{%
\psfig{figure=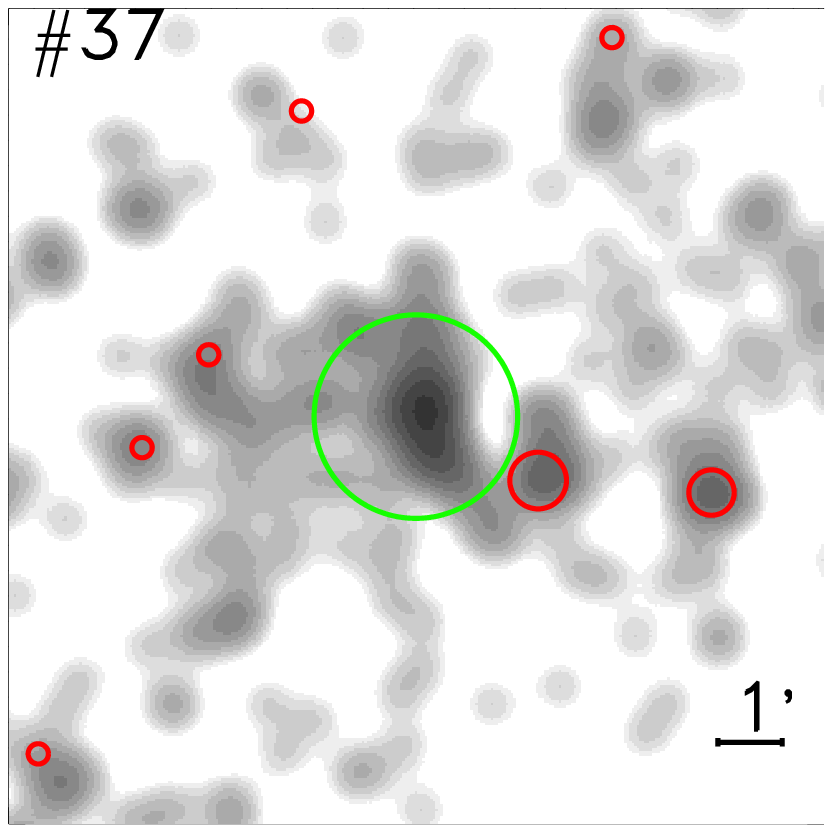,width=2.8truecm,clip=}
\psfig{figure=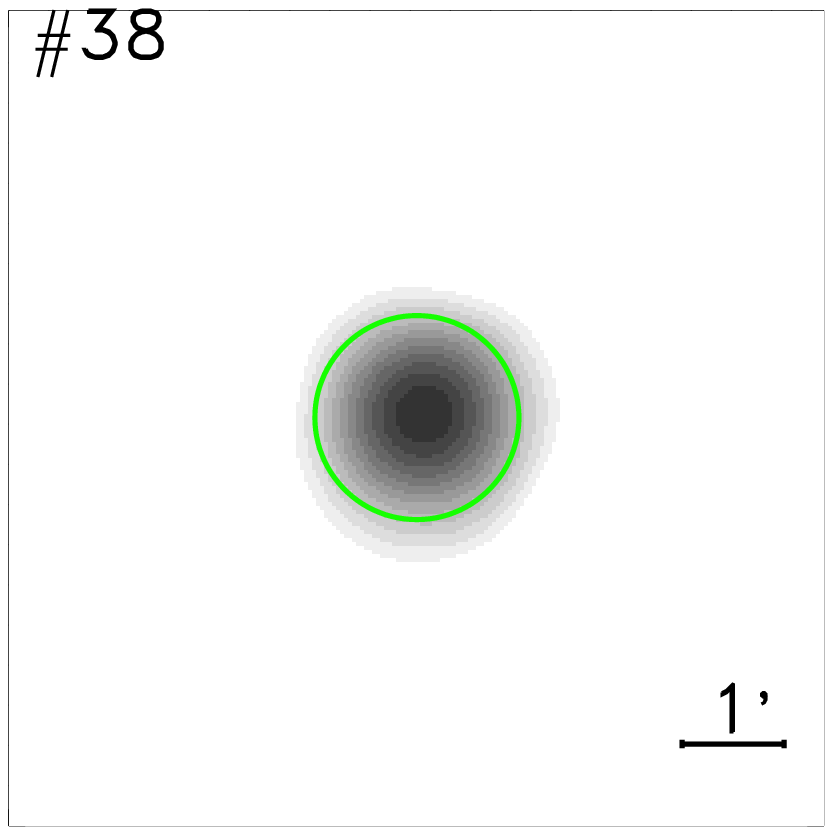,width=2.8truecm,clip=}
\psfig{figure=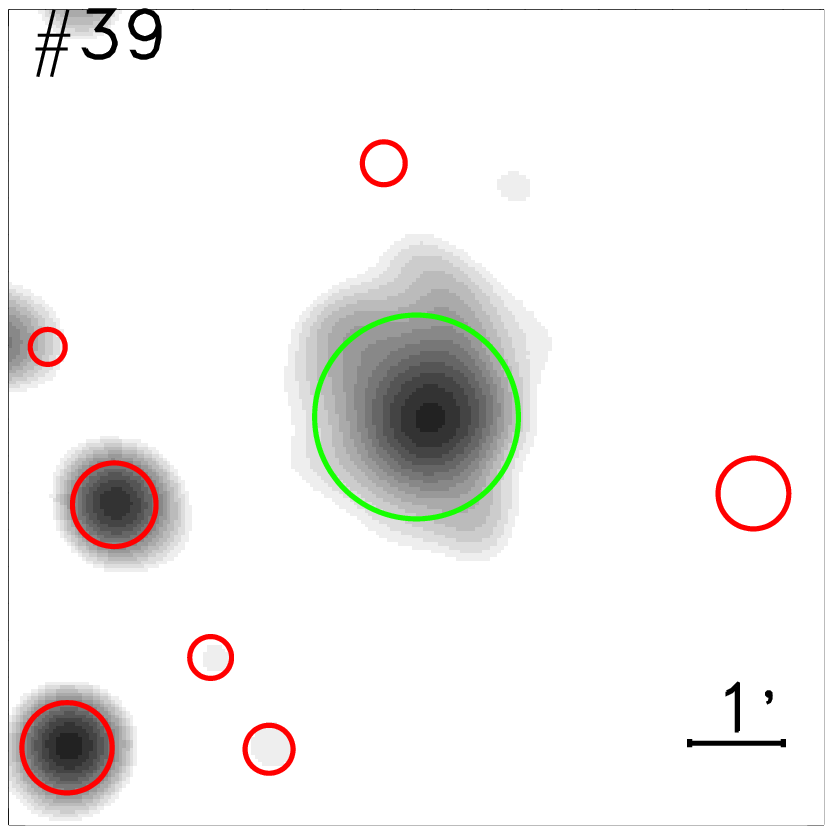,width=2.8truecm,clip=}
\psfig{figure=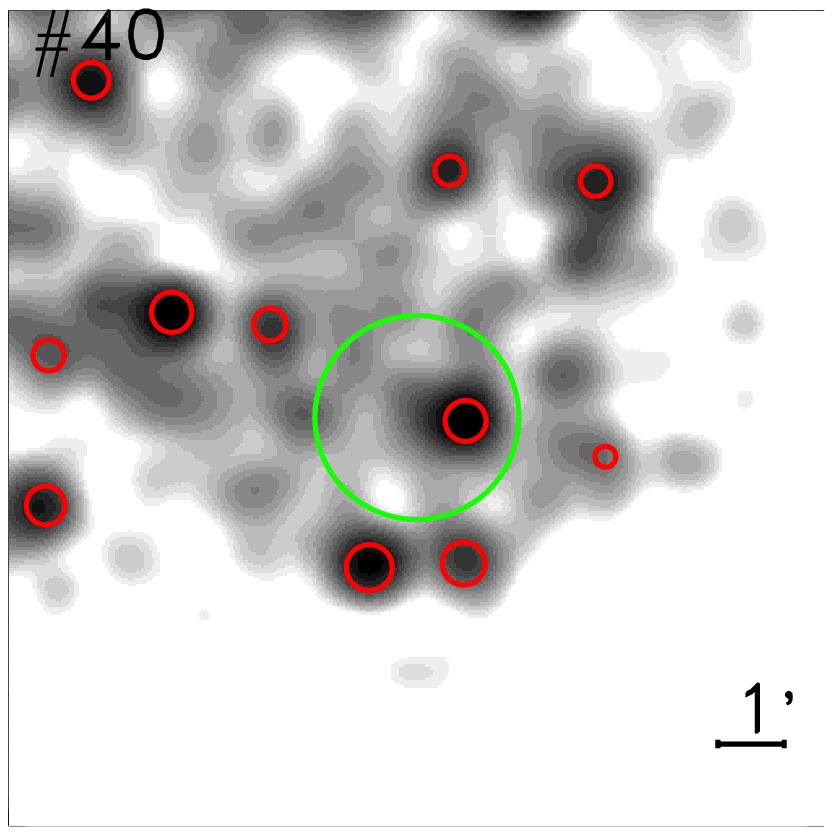,width=2.8truecm,clip=}
\psfig{figure=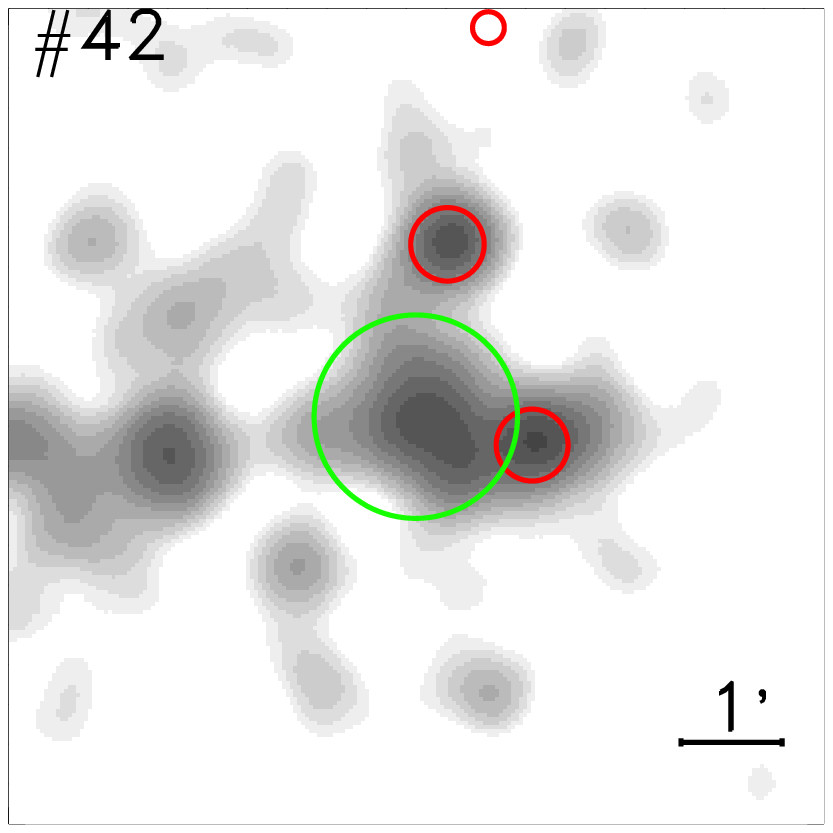,width=2.8truecm,clip=}
\psfig{figure=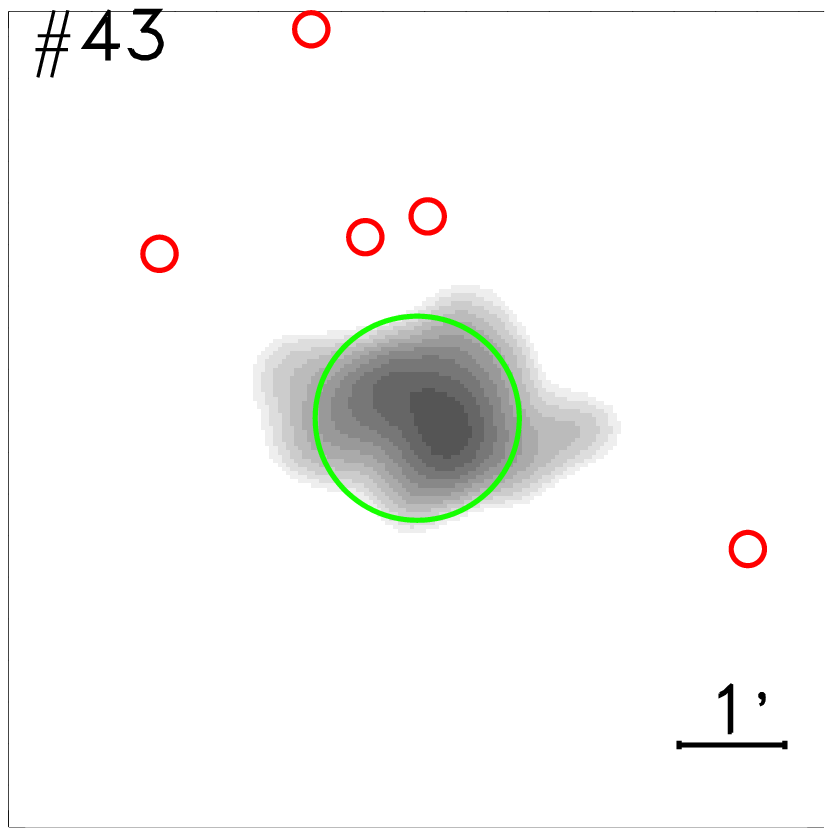,width=2.8truecm,clip=}
}
\caption[h]{Montage of the SWIFT XRT [0.5-2] keV images of 
clusters in our sample, convolved
with a Gaussian kernel with $\sigma=18$ arcsec. 
The ruler indicates 1 arcmin. North is up, east
is to the left. The cluster is marked by a green circle.
Red circles mark other sources masked in our analysis.
}
\end{figure*}

\section{X-ray data}

The X-ray telescope (XRT) on board the Swift satellite (Gehrels et al.
2004) uses a Wolter I mirror set, originally designed for the JET-X
telescope (Citterio et al. 1994), to focus X-rays (0.2-10 keV) onto a
XMM-Newton/EPIC MOS CCD detector (Burrows et al. 2005). The effective area
of the telescope ($\sim$ 120 cm$^2$ at 1.5 keV) is $\sim$ 3.5 smaller than
1 XMM-Newton MOS module. The PSF, similar to XMM, is characterized by a
half-energy-width (HEW) of $\sim$ 18\arcsec  at 1.5 keV (Moretti et al.
2005).

XRT data were reduced using the standard data
reduction procedures as outlined in Moretti et al. (2009).
Two of our clusters, \# 12 and 34, are angularly not far from a 
background gamma-ray-burst. In these cases, we removed the first 
segments of the observations to reduce
the noise associated with the bright gamma-ray-burst.

The total SWIFT XRT integration time on our cluster
sample is 1.4 Ms. 

Figure 4 shows [0.5-2] keV images, 
convolved with a Gaussian kernel with $\sigma=18 $ 
arcsec. The extended X-ray emission of most of them is fairly obvious
in this figure.

To estimate the cluster count rate, we measured
counts (in the [0.5-2] keV band) in the cluster direction, $obstot_i$, 
within a 500 kpc aperture
at the cluster redshift, centered on the revised
cluster center. To estimate the background and its fluctuations,
we measured the counts in a number ($nbox_i$) of regions 
of the same solid angle as the cluster, 
spread over the XRT field-of-view.

\begin{figure}
\centerline{\psfig{figure=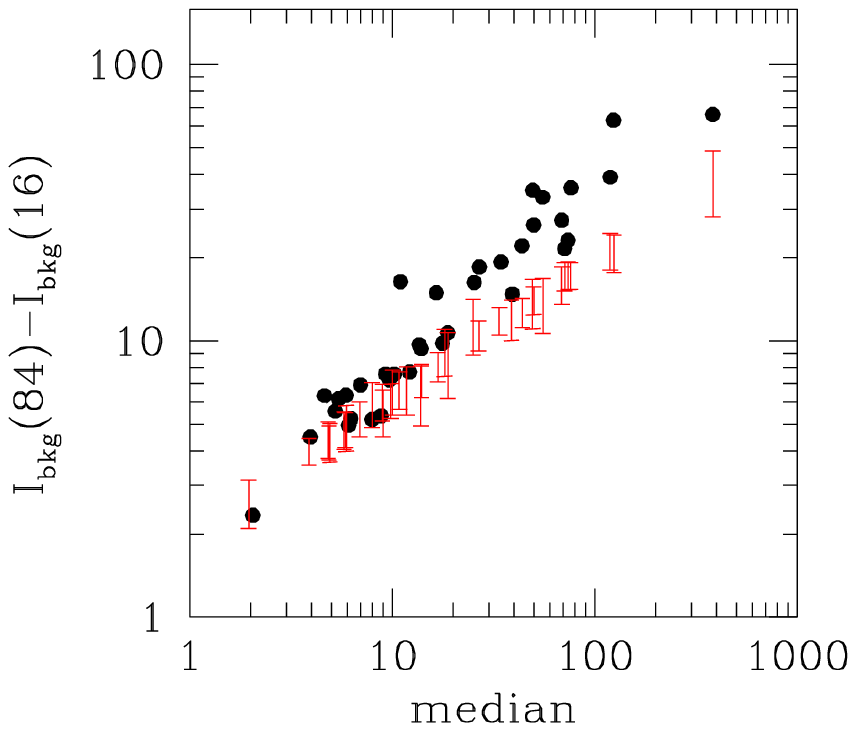,width=6truecm,clip=}}%
\centerline{\psfig{figure=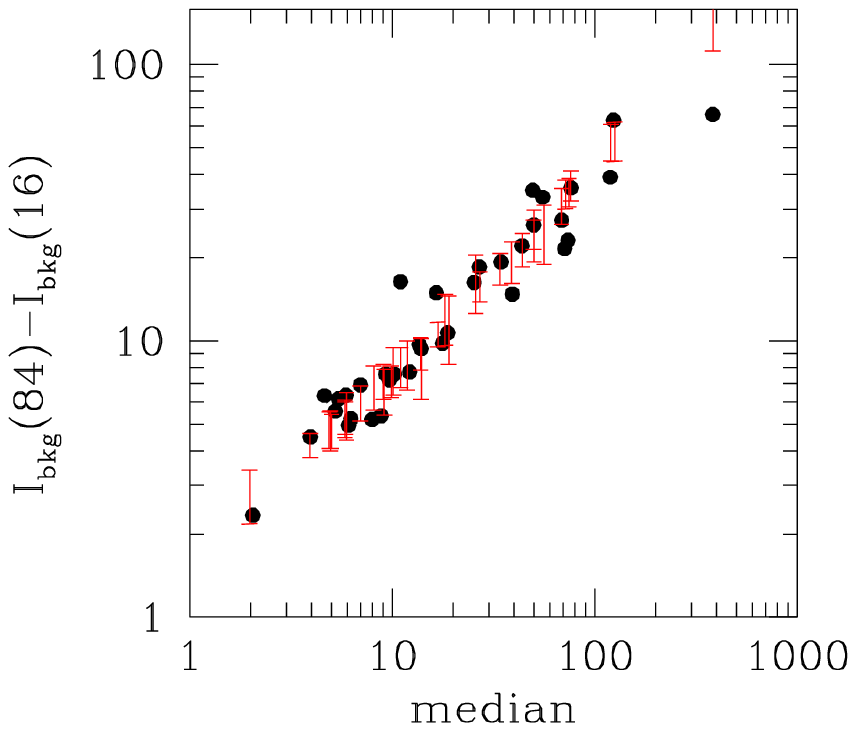,width=6truecm,clip=}}
\caption[h]{Expected (red error bars) and observed (black points)
spread in background values vs median background value. The
upper panel assumes purely Poisson background fluctuations, whereas
the lower panel allow  
a 20 \% scatter in the mean background value. }
\end{figure}

To eliminate the contamination by point sources,
we ran the {\it wavedetect} CIAO task and we masked X-ray point
sources associated to galaxies or optical
point objects. This left us with only the signal coming from the ICM.
We used exposure maps to calculate the effective
exposure time accounting for vignetting, CCD defects, and
excised regions.

To estimate possible over-Poisson fluctuations of background counts,
we computed the 16$^{th}$, 
50$^{th}$ (median) and 84$^{th}$ percentiles of the distribution of
background values and from these the spread, $I_{bkg}(84)-I_{bkg}(16)$, 
plotted in the abscissa of Figure 5. 
We then computed the same percentiles for a sample of
simulated $nbox_i$ background values drawn from a Poisson 
distribution of mean intensity $\mu$. The upper panel of Figure 5 shows that the
observed spread (solid points) is larger than the simulated one
(error bar, showing the 68 \% range of simulated spreads) 
assuming Poisson fluctuations only. Note that the
observed spread is noisy and has a systematic
bias because we measured it
from a finite number of elements. Noise and systematic are both
addressed by our simulation.  The lower panel
shows simulations that better match the observed spread: we
allowed the background to have a  20 \%
Gaussian fluctuation on the top of the Poisson fluctuations.
The agreement between observed
and simulated spread is fairly good, and therefore, we allowed
20 \% over-Poisson background fluctuation throughout. 

Of course, we kept separate data from different
pointings in our calculation  because we are interested in background variations
on cluster angular scales.

To summarize, we found that the $nbox_i$ background values scatter
more than expected if the only source  of background fluctuations
were Poisson i.e. we detected over-Poisson fluctuations of
background counts consistent with a 20 \% amplitude. Some
over-Poisson fluctuation is expected (Moretti et al. 2011).

Modeling this term is important for low surface brightness objects
whose intensity is heavily affected  by a 20 \% background
variation. In particular, our modeling of the over-Poisson
fluctuations of the background is important for the three faintest
clusters, \#9, 36 and 40, which otherwise would have their
luminosity error  underestimated by about 0.2 dex.

Figure 6 shows X-ray counts in the direction of the clusters in
units of the mean background value measured all around them,
together with background errors and 20 \% over-Poisson
fluctuations. All clusters, except \#9, display a significant
excess  of X-ray counts in a 500 kpc aperture. For cluster \#9, we
measured a flux excess higher than expected Poisson fluctuations,
but well consistent with a 20 \% background fluctuation.  
Clusters \# 36 and \# 40 have also low S/N X-ray counts in the 500
kpc aperture, but their detection is secure adopting an optimized
aperture.

\begin{figure}
\centerline{\psfig{figure=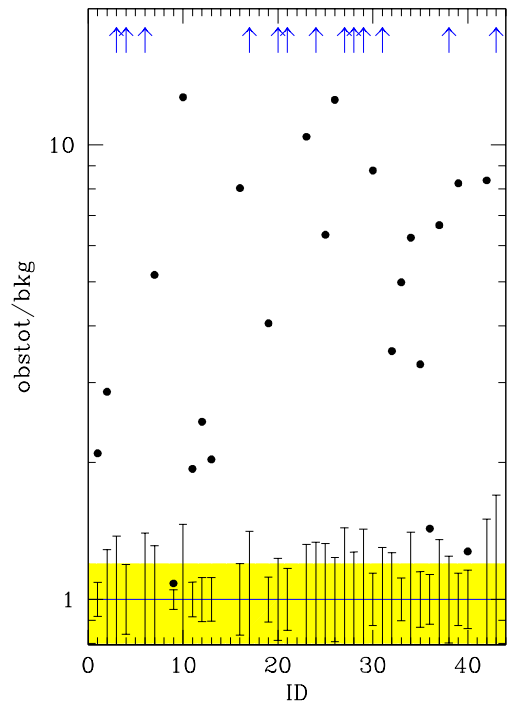,width=7truecm,clip=}}
\caption[h]{X-ray counts in the direction of the clusters (points)
in units of the mean background value measured all around them. 
The error bars indicate heuristic ($\sqrt n$) background Poisson 
fluctations.
The (yellow) shading indicates the 20 \% over-Poisson fluctations.
Thirtenn clusters, indicated by a lower limit, are too bright to fit in this figure.
}
\end{figure}

To ascertain the extension of the X-ray emission,  we calculated
for each cluster the half-power-radius (HPR), defined as the
radius enclosing 50\% of the fluence within a 1 arcmin (25 pixels)
circle radius, which corresponds to the $\sim$ 95\% of the
PSF encircled energy fraction. We assessed the significance of the
extension of each cluster simulating 1000 PSFs with the same
counts, the same spectrum and same off-axis angle. To each
simulated PSF we added a background, accounting for its whole
variance (Poisson and over Poisson). The typical HPR of a point
-like source is 3 pixels (7 arcsec) with a distribution tail that
mostly depends on the signal-to-noise ratio. Fig. 7 shows that for
all the clusters of our sample the HPR lies well beyond the
90$^{th}$ percentile of the HPR PSF simulation distributions,
except \# 40, which barely exceeds it. In this case a significance
near 100 \% is precluded by small number statistics and high
allowed (Poisson plus over-Poisson) fluctuations.

\begin{figure}
\centerline{%
\psfig{figure=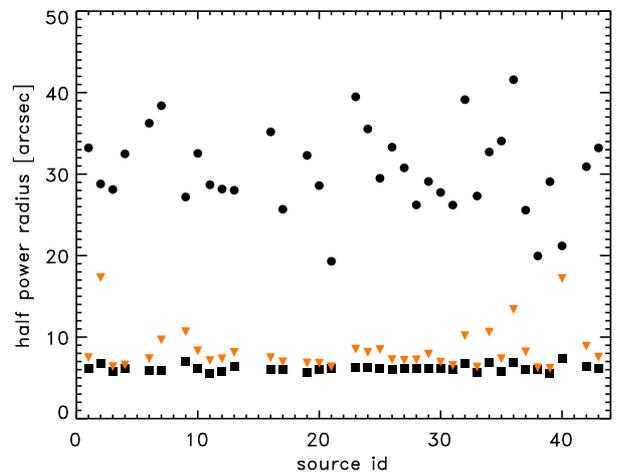,width=9truecm,clip=}
}
\caption[h]{Half power radius of clusters (black circle) 
compared with the $50^{th}$ (square) and $90^{th}$ (orange triangle)
percentile of the distribution of PSFs simulated in the same conditions
(off-axis, spectrum, counts and background).}
\end{figure}

We computed the X-ray count rate of the clusters in our sample
using the fitting model in Appendix A, which  accounts for the
Poisson nature of counts, over-Poisson background fluctuations,
uncertainty on the mean  value of the background, and the existence
of boundaries in the data and parameter space. 
Cluster counts were converted into X-ray luminosities accounting
for the exposure map and assuming a thermal spectrum (APEC) of
$T=1.5$ keV, { 0.3 times the solar value} metallicity, at the
cluster redshift and the Galactic absorption (Kalberla et al.
2005).  We checked that using a $T=3.5$ keV temperature does not
alter our conclusions.
Figure 8 shows the (posterior) probability distribution
of $L_X$ for our clusters. Sharp distributions indicate precisely
determined $L_X$. Note the asymmetry and general
non-Gaussian shape of clusters with lower quality determinations
of $L_X$.

We detail the fitting model in Appendix A in a user-friendly 
way for computing the flux, and/or its  upper
limit, of whatever source. Again, the use of this model is
particularly important for the faintest elements of our catalog,
which are the most interesting cases for the purposes of this
work. The fitting model
returns physically acceptable values in all situations, including 
when observed counts in the cluster direction are lower than the
average measured background, a situation that occurs, for
example, when a faint cluster is on the top of a negative
background fluctuation. Returned uncertainties behave as 
expected: they do not include non-physical (negative) X-ray
luminosities and are large when the X-ray flux is low,  which is
not guaranteed in other approaches (as illustrated in Kraft et al.
1991).

\begin{figure}
\centerline{%
\psfig{figure=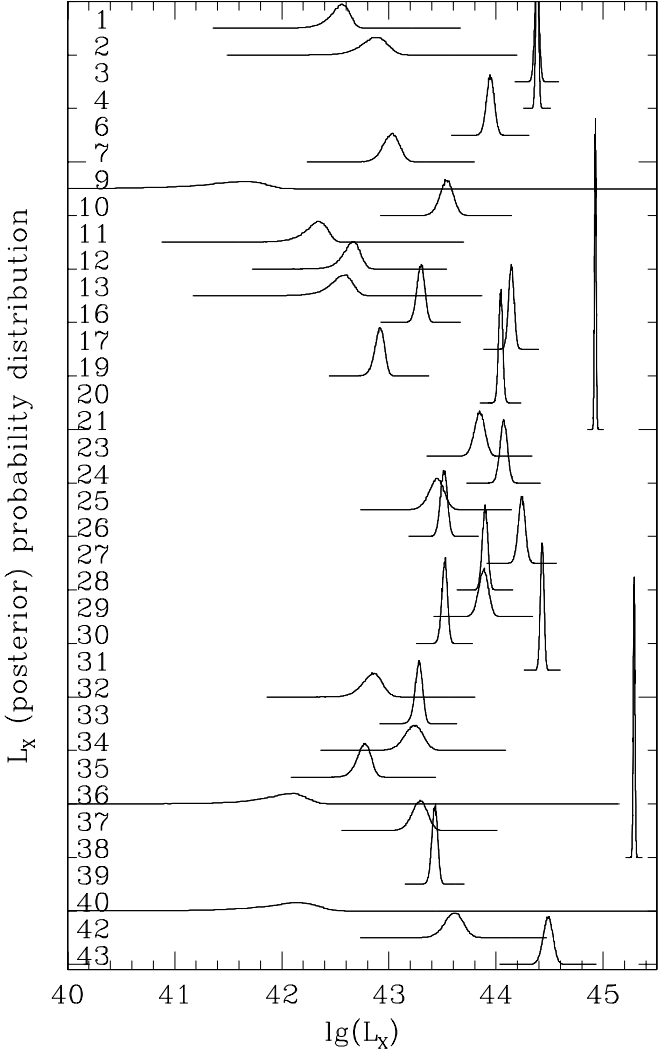,width=9truecm,clip=}
}
\caption[h]{(Posterior) $L_X$ probability distribution for
clusters in our sample. Curves are offset vertically to
improve readibility. Note the non-Gaussian shape (e.g. asymmetry)  
of the several of them, as also indicated by asymetric errors in
Table 2.}
\end{figure}

Table 2 lists cluster id (column 1), Swift exposure time on source
(column 2), total number of photons in the cluster direction,
$obstot_i$ (column 3), and in the background direction,
$obsnbkg_i$ (column 4). The latter is measured in a solid angle
$nbox_i$ times larger.  Column 6 gives the number needed to
convert counts into X-ray luminosity.  Column 7 lists derived X-ray
luminosities and their 68 \% (highest posterior) intervals. 

To summarize, all clusters, except \#9, display a significant extended 
X-ray emission, as shown in Fig 6 \& 7. 
As mentioned, the X-ray emission of cluster \#9 is not secured. The X-ray
detection of 32 clusters out of 33
implies that the 90 \% upper limit fraction of X-ray dark clusters 
is $0.11$. This number 
should be read as pessimistic because cluster \#9 is likely an X-ray emitting system. 
Therefore we can infer that X-ray surveys do not systematically miss a significant population of 
X-ray dark halos, which is an essential assumption for any cosmological use of X-ray cluster surveys such as 
those that will be performed e-Rosita and, possibly, WFXT.

The fraction of color-detected clusters that are real objects is even higher, 0.935  (at 90
\% confidence), because cluster \#9, for which our X-ray data provide no compelling
evidence, is spectroscopically confirmed. In other words, we find that all color-selected
clusters are real (the 90 \% upper limit to spurious detection is $0.065$); this is an
essential and promising piece of information for incoming surveys as DES or EUCLID.

Our finding of a tight upper limit to the fraction
of X-ray dark clusters agrees with the results of Donahue et al.
(2001), but offers a more stringent constraint. While
up to 75 \% of the optically selected clusters in
Donahue et al. (2001) might be dark because
X-ray undetected, our 90 \% upper limit is around
5 to 10 \%. We find a tight upper limit
because our observation strategy has been tailored to avoid
little informative upper limits to the X-ray flux, i.e. values 
brighter than, or comparable to, the mean $L_X$-richness relation.

Our upper limit to the fraction of X-ray dark cluster agrees
with the fraction one can derive from the
X-ray observations of 13 clusters
at much higher redshift, $0.6<z<1.1$ reported in 
Hicks et al. (2008) and Bignamini et al. (2008).
Counting as possibly dark all clusters that do not have
a clear X-ray detection, one in our sample and three in theirs,
the 90 \% upper limit
to the fraction of X-ray dark clusters is 11 \% in our sample, and 42 \%
in theirs. The latter value is higher because
their sample size is small, only 13 systems, and their X-ray upper
limits are little informative.

\input Lx.table.tex

We stress that our conclusion are applicable to color-detected clusters in the local Universe
($0.1<z<0.3$) using a filter pair that brackets the  4000 \AA \  break, of richness
comparable to the clusters listed in the maxBCG catalog  ($>10$ galaxies counted as they do,
or $\ge4$ as we do). As just mentioned, there are indications that the same results may
hold true at higher redshift.

We note that the fraction of spurious detection that we find in the MaxBCG catalog ($<$ 6.5\%)
is consistent with the typical contamination level of  X-ray and SZ catalogs. For example, 1 in 34 of the 
REFLEX (i.e. X-ray) selected clusters have subsequently been discovered to be AGNs (Boheringer et al. 2007);
a similar fraction of objects are expected to be false positives in the 400d survey (Burenin et al. 2007). 
Four out of 21 new cluster candidates identified in the Planck ESZ sample are known not to be single
clusters and are instead double or triple systems from XMM follow-up observations (Aghanim et al. 2011).
Note that both these X-ray and SZ selected samples were subject to attentive scrutiny in the optical prior 
to publication,
whereas our sample of maxBCG clusters were not filtered out by any X-ray data inspection.

\section{$L_X-$richness relation}

As we have said, the 90 \% (pessimistic) upper limit fraction of X-ray dark
clusters is 11\%. We will now to address a finer question, namely whether a significant population
of underluminous clusters exists at all. They may, of course, without being dark (at least in
principle), and may thus have been counted as X-ray emitters in the previous section. Therefore, we 
looked for outliers in the regression between richness and X-ray luminosity.

For this regression, our fitting model assumes a linear relation between (the log of) the
true richness and true X-ray flux (with some intrinsic scatter), but rather than 
these true values, we have noisy  measurements of both richness and X-ray flux, with noise
amplitude different from point to point.   We account for the Poisson nature of counts, the
non-Poisson nature of X-ray flux and richness,   for higher than Poisson fluctuations in the
X--ray  background and covariance for all modeled quantities. This computation requires the
use of the full probability distribution for intervening quantities, not just the  point
estimates of X-ray flux and richness given in Table 1 and 2. The fitting model is fully
described in  Appendix B, where we also give its coding. To our best knowledge, this model
and the one of in Appendix A have never been published before.

\begin{figure}
\centerline{\psfig{figure=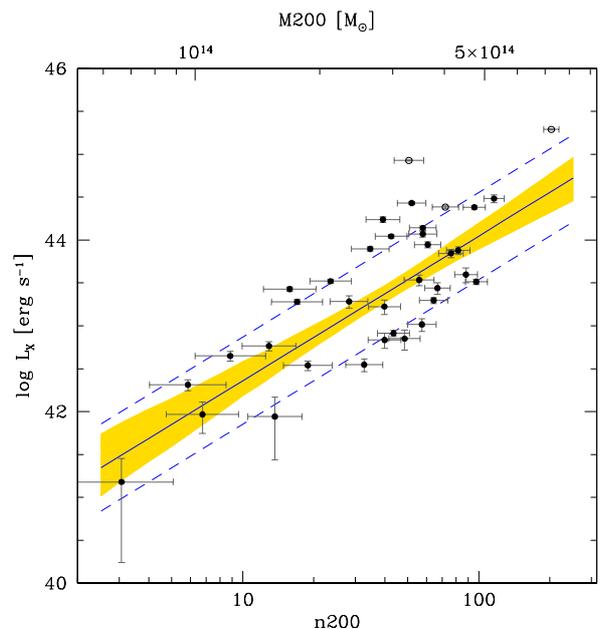,width=8truecm,clip=}}
\caption[h]{X-ray luminosity-richness scaling. 
The solid line indicates the mean fitted regression line of $\log L_X$,
measured within a 500 kpc aperture, on
$\log n200 $,  while the dashed line shows this mean plus or minus the
intrinsic scatter $\sigma_{scat}$. The shaded region marks the 68\% highest
posterior credible interval for the regression. Error bars on the data points
represent observed errors for both variables (computed
following the usual astronomical practice). The distances between the data
and the regression line is partly caused by the measurement error and partly
by the intrinsic scatter. The upper abscissa indicates the cluster mass.
}
\end{figure}
\begin{figure*}
\psfig{figure=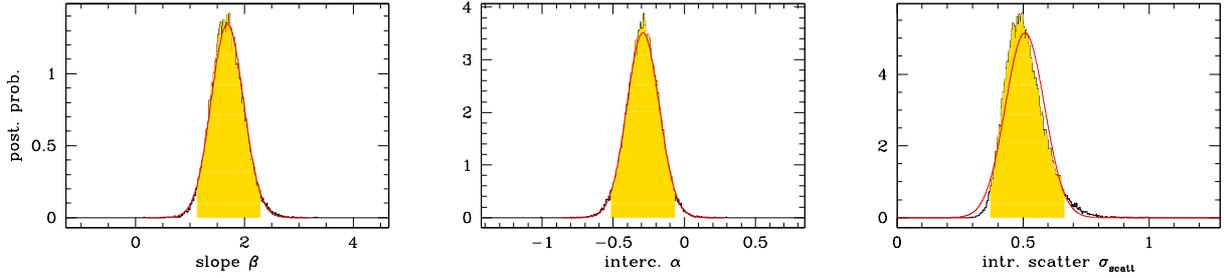,width=17truecm,clip=}
\caption[h]{Posterior probability distribution for the
parameters of the X-ray luminosity-richness scaling.
The black jagged histogram shows the posterior as computed
by MCMC, marginalized over the other parameters. The red curve
is a Gaussian
approximation of it.  The shaded (yellow) range shows
the 95 \% highest posterior credible interval. 
The jagged nature of the histogram is caused
by the finite sampling of the posterior. 
}
\end{figure*}

\begin{figure}
\centerline{\psfig{figure=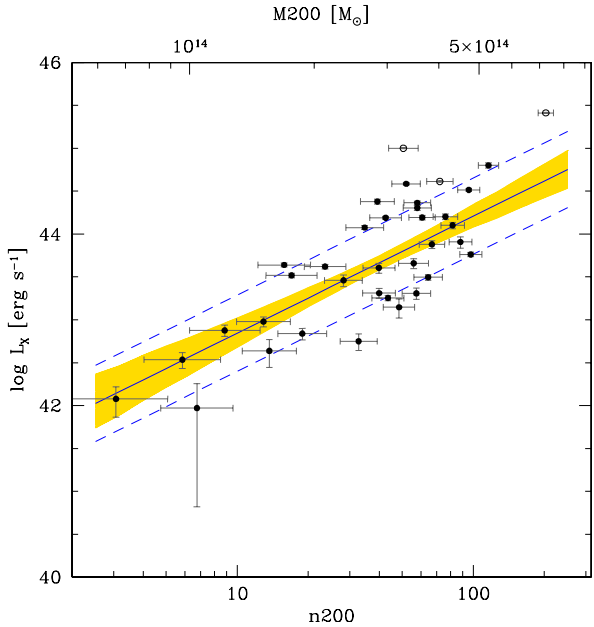,width=8truecm,clip=}}
\caption[h]{As Fig 9, but for an 
$L_X$ measured in an aperture of 1070 kpc. 
}
\end{figure}

Using the fitting model, we found for our sample of 33 color-selected clusters

\begin{equation}
lgL_X = (1.69\pm0.30)  \ (\log n200 -1.8) +43.71\pm0.11 \quad .
\end{equation}

Figure 9 shows the scaling between richness and X-ray luminosity, observed data, the mean scaling 
(solid line) and its 68\% uncertainty (shaded yellow region) and the mean intrinsic scatter 
(dashed lines) around the mean relation.  The $1 \sigma_{scatt}$  band is not expected to contain
68\% of the data points because of the measurement errors. All points are, however, 
within twice the intrinsic scatter. The upper abscissa also gives the cluster mass.

Figure 10 shows the posterior probability distribution of the intercept, slope, and intrinsic
scatter $\sigma_{scat}$. These probability distributions are reasonably well approximated by 
Gaussians. The intrinsic $L_X$ scatter at a given richness, $\sigma_{scat}=\sigma_{lgL_X|\log n200}$, is very 
large, $0.51\pm0.08$ dex. In other terms, a whole 1 dex in $L_X$ is needed to bracket 68 \% of
clusters of a given richness. 

Figure 9 also shows that the three (out of three) X-ray selected clusters (for this reason
not fitted) are much brighter than the mean regression, as expected for X-ray selected
objects. Figure 9 shows that the data of cluster \# 9 (the object with the lowest $n200$ and
an X-ray flux excess higher than the Possion fluctuation,  but still consistent with a 20 \%
background fluctuation) are also compatible with the X-ray luminosity expected  for its
richness.

Figure 9 allows us to address the question mentioned at the start of this section, whether there is a
population of clusters with much lower X-ray luminosity at a given richness or mass, i.e. underluminous. None
of them has been found in our sample, because no point is much  off from the regression, the farthest one
being  about 1.5 times the intrinsic scatter. Indeed, the current data allow us to set an upper limit to the
fraction of underluminous clusters. Because no outlier is present in a sample of 33, this sets a 90 \% upper
limit of 0.065.

We used a 500 kpc aperture as a good compromise between the physical dimensions of the cluster and  the
typical apparent size in our observation. To test the robustness of our results we also measured the
relation using an aperture of 1.07 Mpc (adopted by Rykoff et al. 2008). We found

\begin{equation}
lgL_X = (1.36\pm0.26) \ (\log n200 -1.8) +43.93\pm0.10 
\end{equation}

and an intrinsic scatter of $0.44\pm0.07$. Fig 11 shows the scaling between richness and X-ray luminosity
with this larger aperture. These parameters are consistent with those derived using the small
aperture. If anything, the intercept is slightly larger than using a smaller aperture because there is some cluster
flux outside $0.5$ Mpc. As in the case of the smaller aperture, there are no outliers and no cluster 
qualifies itself as underluminous.

Slope and intercept agree with the values reported in Rykoff et al. (2008).
Our intrinsic scatter, which formally agreement, is larger than the total scatter they
report, even though we expect
the contrary because their (total) scatter does not account for point-source contamination, 
for miscentering (30 \% of clusters with wrong coordinates), redshift dependence 
of their definition of richness, and richness 
errors\footnote{Rozo et al. (2011) prefer to use the term ''intrinsic scatter" to indicate
richness errors.}. We checked
to find a larger scatter also adopting the MaxBCG $n200$ definition. 
We emphasize that while these observed differences are within the errors, 
our work gives the first robust measurement of the {\it intrinsic} 
scatter of $L_X$ at a given richness, previous attemps 
(e.g. Rykoff et al. 2008) do not have
completely removed all observationally-related effects from its
estimate.

\section{Do we know any underluminous clusters?}

We found no underluminous clusters. However, other works 
(e.g. Bignamini et al. 2008, Hicks et al. 2008, Castellano et al. 2011,
Dietrich et al. 2009, Balogh et al. 2011) do. What is the reason for this? 
To claim that a cluster is underluminous, it is critical 

a) to account for the intrinsic scatter. If none is assumed,
a cluster is claimed to be underluminous when instead it is normal, i.e. just one (true)
$\sigma_{intr}$ below from the mean relation. Once the
intrinsic scatter is allowed, the putative
Dietrich et al. (2009) X-ray underluminous cluster becomes
a normal cluster.

b) to use a non-biased relation as reference to determine the
underluminous nature of a putative cluster. Indeed, if a reference
relation is (mis)taken biased-high, one may wrongly classify 
normal clusters as underluminous. The reference relation may be easily
biased as high if the comparison sample is formed by X-ray selected
clusters or by an uncontrolled sample, as is now well known from the
literature (e.g. for the $L_X-T$ relation: 
Pacaud et al. 2007; Stanek et al. 2006;
Nord et al. 2008; Andreon et al. 2011; Andreon \& Hurn 2011). 
The bias occurs when 
the probability that a cluster is included in the sample
depends on its own $L_X$. This is the case for an X-ray selected
sample, but also for every sample for which there is a
selection based on (individual) cluster luminosity, flux or counts, 
(e.g. at least $n$ photons for a temperature measurement,  with
$n$ often in the range 200-1000).
Indeed, in an X-ray selected sample, the bias
comes from the larger Universe volume over which a cluster 
brighter-than-average at a given richness can be seen compared
to fainter-than-average clusters. Therefore, in an X-ray selected sample, at a given
richness the upper half of the $L_X$ distribution will be more
populated than the lower half, biasing high the mean, and
underestimating the dispersion (if the effect is not accounted for).
A similar bias is also likely present in cluster samples
assembled from pointed observations of X--ray selected clusters
like the ones built by, amongst others, Ettori et al. (2004) 
and Branchesi et al. (2007).
On the contrary, a purely optically (or color, as in this
work) selected sample of clusters does not
introduce any bias in the $L_X$-richness relation,
because the average $L_X$ at a given richness will not
be biased high (or low).

Bignamini et al. (2008), Hicks et al. (2008) conclude that 
color-selected clusters are underluminous. Castellano et al. (2011)
claim the existence of an underluminous cluster.
However, their
claim is based on the comparion with a biased-high mean
$L_X-T$. 

Finally, Balogh et al. (2011) asses the X-ray
properties of their sample using
Chandra or XMM observations for all but two clusters (both
undetected in X-ray), and claim the
existence of five X-ray underluminous clusters in a sample of 18 of mass
in the range considered in our sample.
However, a) their X-ray upper 
limits are all equal irrespective of the cluster redshift,
exposure time, or X-ray telescope used
(Rosat vs Chandra or XMM); b) four out five of them are 
within 2 sigma (1 dex) of the mean $L_X$-richness relation, not
enough to call them underluminous (outlier); c) the
authors note that three of the five underluminous clusters,
objects 13, 17 and 18,
are possibly fake objects (chance projections), not truly existing
clusters.

Older claims about the existence of underluminous clusters
are rebutted in Andreon et al. (2009). That paper shows that underestimated
errors may incorrectly lead to classify a cluster as underluminous 
even when it agrees with the mean relation, for example when it
is, say, 3 (wrong) $\sigma$ below the mean relation.

\section{Discussion}

The analysis of our sample of 33 clusters at $0.1<z<0.3$ 
and our revision of the recent literature
results presented in Sec 6, joined to our revision of older
works in Andreon et al. (2009), 
confirms that X-ray underluminous clusters are rare enough that we 
are still looking for an example. 
Some scenarios of cluster formation predict the existence of
underluminous clusters, objects in which the gas has been 
expelled (e.g. Bower et al. 2008, McCarthy et al. 2011).
As we found none in a sample of 33, our
90 \% upper limit to this type of objects is $0.065$.

This work, which is based on a color-selected sample of clusters,
does not address the fraction of clusters without a red sequence.
However, past works have shown the absence of X-ray selected clusters without
red galaxies, for example 54 out 54 X-ray selected clusters studied in Andreon
\& Hurn (2010) have red galaxies and all 32 clusters in Garilli et al. (1996)
and in Puddu et al. (2001) of the Einstein Medium Sensitivity Surveys have a red
sequence. Overall, the general picture that emerges from this work, when joined
to the absence of X-ray selected  clusters without red galaxies, is that
outliers in the dark (X-ray or low richness) side are quite rare and that the
X-ray and color selection sample the same population of objects. This
conclusion is supported indeed at much higher redshift by the smaller sample
analyzed in Bignamini et al. (2008) and Hicks et al. (2008), in which no
believable outliers from the mean relation is found, and by the detection of a
red sequence (Andreon et al. 2004; 2005) in all X-ray selected clusters of the
XMM-LSS survey (Pierre et al. 2004).

\section{Summary and conclusions}

We studied the X-ray properties of a color-selected sample of
clusters at $0.1<z<0.3$ and we critically discussed previous works claiming
the existence of underluminous clusters.

Two important guidelines have been strictly followed in the sample
selection.  First, the sample has been selected to have sufficiently deep X-ray
observations to probe their X luminosity down to very faint
values and, second, at the same time we did not use any criterium that depends directly
or indirectly on the X luminosity, at a given richness, 
of the single objects.
Our sample consist of 33 clusters that fall in sky
regions where deep Swift XRT X-ray observations are available. 

Using SDSS
data, we refined the cluster centers and richnesses and we
estimated cluster masses using richness as mass proxy.
These clusters have
masses in the range between 5 $10^{13}$ and  8 $10^{14}$ solar masses. This
allowed an unbiased measure of the $L_X$-n200
relation for a small, but representative, sample of galaxy clusters.

Using 1.4 Ms Swift XRT data, we measured the X-ray luminosity within an
aperture of 500 kpc. In these calculations, we accounted for terms usually
neglected, such as over-Poisson fluctuations of X-ray background counts, which
turned out to be on the order of 20 \%, cluster miscentering  (i.e. that the
cluster center is in 30 \% of the cases at a sky location different from
what  is listed in the catalog), the positively defined nature of measured
quantities (richness and X-ray luminosity), etc.

Thirty-two out of our 33 color-selected clusters are obvious X-ray detections. The
remaining cluster shows an X-ray excess in the cluster direction compatible
with a possible background fluctuation but also with the expected X-ray
luminosity of a cluster of the same richness.  
Therefore, the fraction of X-ray dark clusters (if any of them exist)
is low: 11\% (at 90 \% confidence level).

Since t32 out 33
color-detected clusters are X-ray emitting, then at
least 89 \% of  color-detected clusters are real objects with a potential
well deep enough to heat and retain an intracluster medium. 
Because the system with suggestive, but not compelling evidence of an X-ray
emission is spectroscopically confirmed, the fraction of false
positive in color-selected searches has most probably an upper limit of
6.5 \%. The low contamination of color-selected clusters  is a
requirement for the use of color-selected  clusters for cosmological aims
and this work directly shows that this requirement is fulfilled in the mass
and redshift ranges considered here.

The quite strict upper limit  to the fraction of X-ray dark or underluminous
clusters, 6.5 to 11 \%, depending on the status of our system without 
compelling evidence of an X-ray emission, also justifies the widespread
use of X-ray selected clusters (e.g. Pacaud et al. 2007), in the sense that
X-ray surveys do not systematically miss halos with (red) galaxies inside
them. Broadly speaking, X-ray and color (galaxy) cluster searches detect the
same population: no X-ray dark cluster is found (in this work and in
our revision of other works), and no
X-ray selected cluster is found (in other works) not to have red galaxies
(of course, in the mass range and in the portion of the Universe volume
explored by the considered data).

X-ray luminosity, measured within a 500 (1070) kpc aperture,  scales
with richness with a proportional factor  1.69 (1.36), with a noticeable
scatter, $0.51\pm0.07$ ($0.44\pm0.07$) dex. The intrinsic scatter is
compatible, but  larger, than previously reported scatters. 
Nevertheless, we emphasize that previous attemps (e.g. Rykoff et al. 2008)
do not have completely removed all observational-related effects from the
scatter estimate, and thus are not quoting a measurement of scatter
entirely related to the object under study, i.e. intrinsic to clusters, but
a mix of intrinsic scatter and observer-related effects (such as having
no flagged point-sources or having centered the X-ray aperture away
from the cluster).

Finally, we found that the observed fraction of miscentered clusters is 
$0.28\pm0.07$. This parameter and its uncertainty are required to 
estimate cosmological parameters or to perform cosmological
forecasts using richness as mass proxy. 

Our results are very promising for cosmological estimates based on
galaxy-detected clusters at least for the redshift and mass ranges
considered in this work: surveys as DES or EUCLID will image a large
part of the sky, returning  color-selected clusters with the same low
contamination of X-ray selected clusters (as shown in this work), with a
mass proxy of equal quality (Andreon \& Hurn 2010), but with an at least 10
times larger sample (Andreon \& Hurn 2010). Galaxy-detected clusters are
also promising when compared to current SZ-detected clusters:  current SZ
surveys return  one hundred to one thousand fewer clusters than optical
searches  (e.g. one cluster per 8 to 19 deg$^2$,  Vanderlinde et al. 2010;
Marriage et al. 2011), with a mass proxy that has an observationally
determined scatter of about 0.5 dex (Rines et al. 2010), i.e. much worser
than $n200$ performances (0.3 dex, Andreon \& Hurn 2010). Forecasts specifically
for EUCLID will be presented in Trotta et al. (in preparation).

\begin{acknowledgements}
We acknowledge financial contribution from the agreement 
ASI-INAF I/009/10/0 and ASI-INAF I/011/07/0.
For the standard SDSS acknowledgement see: 
http://www.sdss.org/dr6/coverage/credits.html
\end{acknowledgements}

\begin{appendix}

\section{Model for the X-ray flux, accounting for over-Poisson background 
fluctuations}

The aim of this section is to present a Bayesian analysis of the
X-ray luminosity fitting model.
In particular, we wish to acknowledge the uncertainty in all 
measurements, including the background estimation.

Because of errors, observed and true values are not identical.
We call $nclus_i$ and $nbkg_i$  the true cluster
and the true background counts in the studied solid angles.
We measured the number of photons in both cluster and background
regions, $obstot_i$ and $obsbkg_i$ respectively, for each of our 36
clusters (i.e. for $i=1,\ldots,36$). The background solid
angle is $nbox_i$ times larger than the cluster solid angle.
We assume a Poisson likelihood for both and that
all measurements are conditionally independent.

\begin{eqnarray}
obstot_i &\sim& \mathcal{P}(nclus_i+nbkgind_i/nbox_i) \\
obsbkg_i &\sim& \mathcal{P}(nbkg_i) \quad,
\end{eqnarray}

where the symbols $\sim$ reads ``is distributed as" and 
and $\mathcal{P}$ stands for the Poisson
distribution.

$nbkgind_i$ is allowed to fluctuate by 10 \% around the
global background value, so that 
the predicted scatter of background values
matches the observed one inside each
XRT field: 

\begin{eqnarray}
nbkgind_i &\sim& \mathcal{\log N}(ln(nbkg),0.2^2)  \quad ,
\end{eqnarray}

where the symbol $\mathcal{\log N}$ stands for the lognormal distribution. 

We assume uniform priors on cluster and  background
counts, zero-ed to un-physical values:

\begin{eqnarray}
nclus_i &\sim& \mathcal{U}(0,\infty) \\
nbkg_i &\sim& \mathcal{U}(0,\infty) \quad .
\end{eqnarray}

Finally, cluster net counts, $nclus_i$, 
are converted into X-ray luminosities as usual: 

\begin{eqnarray}
lgLx_i \leftarrow log(nclus_i) +C_i \,
\end{eqnarray}

where the arrow symbol reads ``take the value of", and
$C_i$ is the usual conversion from counts to $L_x$.

Eq A1 to A6 find an almost literal translation in JAGS (Plummer 2008),
Poisson, normal, lognormal and uniform distributions become
{\texttt{dpois, dnorm, dlnorm, dunif}}, respectively. 
JAGS, following BUGS (Spiegelhalter et al. 1995), uses 
precisions, $prec = 1/\sigma^2$, in place of variances $\sigma^2$. 
Furthermore, it uses neperian logarithms, instead of decimal ones.

This model (set of equations) reads in JAGS:

\begin{verbatim}
model 
{
for (i in 1:length(obstot)) {
obstot[i] ~ dpois(nclus[i]+nbkgind[i]/nbox[i])
nbkgind[i] ~ dlnorm(log(nbkg[i]),1/0.2/0.2)
obsbkg[i] ~ dpois(nbkg[i])
nbkg[i] ~ dunif(1,1.0E+7) 
nclus[i] ~ dunif(0,1.0E+7)         
# optional, JAGS is not needed to do it
lgLx[i] <- log(nclus[i])/2.30258 +C[i]      
}
}
\end{verbatim}

This model (and code) gives the posterior distribution of the
cluster X-ray luminosity, given the observed values of cluster
and background counts. Data, posterior mean and (highest posterior)
68 \% intervals of the X-ray luminosity are listed in Table 2.

Note that a different prior for cluster and background
counts may be more appropriate and valuable in other contexts.
In Appendix B
we adopt the prior inherited from the cluster richness 
for the cluster signal.

\section{Model for the X-ray luminosity vs richness/mass}

The aim of this section is to present a Bayesian analysis of the
X-ray luminosity-richness fitting model.
In particular, we wish to acknowledge the uncertainty in all 
measurements, including background estimation. Basically,
our model regresses two quantities, each one given by the difference
of two Poisson deviates (photons or galaxy counts). We allow
the existence of an intrinsic scatter between regressed quantities,
and higher than Poisson fluctuations of the X--ray background. 
In the statistics literature, such a model is know as an 
``errors-in-variables regression'' (Dellaportas \& Stephens, 1995).
Our model is an extension of the model in Andreon \& Hurn (2010),
accounting for the different nature of one of the modeled quantities
(X-ray luminosity instead of mass) and for the presence of over-Poisson
fluctuations.

First of all, because of errors, observed and true values
are not identically equal.
The variables $n200_i$ and $ngalbkg_i$ represent the true richness 
and the true background galaxy counts in the studied solid angles.
We measured the number of galaxies in both cluster and control field
regions, $obsgaltot_i$ and $obsgalbkg_i$ respectively, for each of our 33
clusters (i.e. for $i=1,\ldots,33$).
We assumed a Poisson likelihood for both and that
all measurements are conditionally independent.
The ratio between the cluster and control field solid angles,
$Cgal_i$, is exactly known. In formulae:
\begin{eqnarray}
obsgalbkg_i &\sim& \mathcal{P}(ngalbkg_i) \\
obsgaltot_i &\sim& \mathcal{P}(ngalbkg_i/Cgal_i+n200_i) \quad ,
\end{eqnarray}

For the X-ray photons a similar construct holds, as detailed in the
section above, with eq A4 removed (the prior on $L_X$ is inherited from $n200$ and
$\sigma_{scatt}$ ones), and eq A6 replaced by

\begin{eqnarray}
nclus_i \leftarrow 10^{lgLx_i-C_i} \quad .
\end{eqnarray}

We assume a linear relation between the unobserved $L_X$ and $n200$ on the
log scale, with intercept $\alpha+44.0$, slope $\beta$ and
intrinsic scatter $\sigma_{scat}$:

\begin{equation}
lgLx_i \sim \mathcal{N}(\alpha+44+\beta (\log(n200_i)-1.8), \sigma_{scat}^2) \quad .
\end{equation}

Note that $\log(n200)$ is centered at an average value of 1.8 and
$\alpha$ is centred at 44.0, purely for computational advantages in
the MCMC algorithm used to fit the model (it speeds up
convergence, improves chain mixing, etc.), and that 
the relation is between true values, not between observed values.

The priors on the slope and the intercept of the regression line in
Equation B4 are taken to be quite flat,
a zero mean Gaussian with very large variance for $\alpha$ and a
Students $t$ distribution with 1 degree of freedom for $\beta$.
The latter choice is made to avoid that properties of galaxy clusters 
depend on humans rules to measure angles (from the x axis anticlockwise
or from the y axis clockwise).
This agrees with the model choices in Andreon (2006 and later works)
but differs from most other works. Our $t$ distribution on $\beta$ is 
mathematically equivalent to a uniform prior on the angle $b$.

\begin{eqnarray}
\alpha &\sim& \mathcal{N}(0.0,10^4) \\
\beta &\sim& t_1 \quad .
\label{eqn:eqn11}
\end{eqnarray}

Finally, we need to specify the prior for the intrinsic scatter,
$\sigma_{scat}$, which is positively defined. 
Following Andreon \& Hurn (2010) and Andreon (2010), we impose
a quite weak prior information: a Gamma distribution on $1/\sigma_{scat}^2$,

\begin{eqnarray}
1/\sigma_{scat}^2 &\sim& \Gamma(\epsilon,\epsilon)  \quad ,
\end{eqnarray}
with $\epsilon$ taken to be a very small number.

In JAGS, our model reads

\begin{verbatim}
model
{
intrscat <- 1/sqrt(prec.intrscat)
prec.intrscat ~ dgamma(1.0E-5,1.0E-5)
alpha ~ dnorm(0.0,1.0E-4)
beta ~ dt(0,1,1)
for (i in 1:length(obstot)) {
  # modelling X-ray photons 
  obstot[i] ~ dpois(nclus[i]+nbkgind[i]/nbox[i])
  nbkgind[i] ~ dlnorm(log(nbkg[i]),1/0.2/0.2)
  obsbkg[i] ~ dpois(nbkg[i])
  nbkg[i] ~ dunif(0,10000)
  # convert nclus in Lx
  nclus[i] <- exp(2.30258*(lgLx[i]-C[i]))
  # modelling galaxy counts
  # n200 term
  obsgalbkg[i] ~ dpois(ngalbkg[i])
  obsgaltot[i] ~ dpois(ngalbkg[i]/Cgal[i]+n200[i]) 
  n200[i] ~ dunif(1,3000)
  ngalbkg[i] ~ dunif(0,3000)
  # modeling Lx -n200 relation 
  z[i] <- alpha+44+beta*(log(n200[i])/2.30258-1.8)
  lgLx[i] ~ dnorm(z[i], prec.intrscat)
  }
}
\end{verbatim}

\end{appendix}

\end{document}

%% file: table2.tex
%
\begin{table*}
\caption{Observed galaxy counts, solid angle ratios, and cluster masses.
}
\begin{tabular}{r r r r r r r l l l}
\hline
ID & ra & dec & $obsn(<1.43)$ & $obsgaltot$ & $obsgalbkg$ & $Cgal$ & $n200$ & $lgM200$ & other IDs\\ 
& \multispan{2}{\hfill (J2000) \hfill } \\
%
(1) & (2) & (3) & (4) & (5) & (6) & (7) & (8) & (9) & (10)\\
\hline
 1 & 125.171 & 31.953 & 29.3 & 25 & 61 & 9.95	   & $ 20^{+5}_{-5} $	   & $14.28\pm 0.29$   \\  		      
 2 & 139.148 & 63.803 & 50.6 & 60 & 29 & 2.52	   & $ 49^{+7}_{-8} $	   & $14.50\pm 0.28$   \\  		     
 3 & 145.511 & 8.959  & 83.8 & 115 & 73 & 3.78     & $ 96^{+11}_{-11} $    & $14.67\pm 0.29$   & Abell 854 (1.3') \\  
 4* & 152.199 & 11.799& 70.1 & 95 & 142 & 6.22     & $ 73^{+8}_{-11} $     & $14.60\pm 0.28$   & MS1006+1202 \\       
 6 & 154.152 & 24.801 & 60.2 & 68 & 29 & 4.00	   & $ 61^{+8}_{-8} $	   & $14.55\pm 0.28$   & Abell 964 (0.2') \\  
 7 & 156.988 & 10.579 & 57.4 & 68 & 18 & 1.70	   & $ 58^{+7}_{-10} $     & $14.54\pm 0.28$   \\  		      
 9 & 177.532 & 57.151 & 6.7 & 4 & 19 & 20.33	   & $ 4^{+1}_{-3} $	   & $13.86\pm 0.37$   \\  		      
10 & 177.950 & 37.258  & 55.1 & 72 & 62 & 3.85     & $ 57^{+8}_{-10} $     & $14.55\pm 0.28$   \\  		      
11 & 183.997 & 35.717  & 7.2 & 7 & 39 & 34.20	   & $ 7^{+2}_{-3} $	   & $14.01\pm 0.33$   \\  		      
12 & 184.066 & 35.520  & 16.4 & 13 & 120 & 28.95   & $ 10^{+3}_{-4} $	   & $14.10\pm 0.32$   \\  		      
13 & 184.117 & 35.639  & 38.7 & 41 & 104 & 12.52   & $ 33^{+5}_{-8} $	   & $14.41\pm 0.28$   &Abell 1738 (0.6')\\   
16 & 201.297 & 57.600  & 63.2 & 81 & 29 & 1.75     & $ 65^{+9}_{-10} $     & $14.57\pm 0.28$   & Abell 1744 (1.4')\\  
17 & 201.457 & 59.330  & 58.0 & 64 & 22 & 3.62     & $ 59^{+8}_{-9} $	   & $14.56\pm 0.28$   \\  		      
19 & 211.869 & 27.821  & 49.7 & 53 & 45 & 4.73     & $ 44^{+6}_{-9} $	   & $14.48\pm 0.29$   & Abell 1861 (0.6')\\  
20 & 217.778 & 25.634  & 45.9 & 48 & 11 & 2.01     & $ 43^{+7}_{-8} $	   & $14.47\pm 0.28$   \\  		      
21* & 224.313 & 22.342 & 51.1 & 60 & 86 & 9.07     & $ 51^{+7}_{-9} $	   & $14.52\pm 0.29$   & MS1455.0+232\\       
23 & 228.214 & 14.318  & 69.9 & 89 & 46 & 3.60     & $ 77^{+10}_{-9} $     & $14.61\pm 0.29$   & Abell 2044 (0.3')\\  
24 & 229.075 & 0.089   & 58.0 & 67 & 21 & 2.31     & $ 58^{+7}_{-9} $	   & $14.55\pm 0.28$   & Abell 2050 (1.1')\\  
25 & 233.265 & -0.771  & 61.5 & 76 & 32 & 3.48     & $ 67^{+8}_{-10} $     & $14.59\pm 0.28$   \\  		      
26 & 233.840 & 37.396  & 83.6 & 114 & 33 & 2.01    & $ 98^{+12}_{-10} $    & $14.68\pm 0.29$   \\  		      
27 & 245.254 & 25.772  & 43.9 & 48 & 65 & 7.41     & $ 40^{+7}_{-8} $	   & $14.45\pm 0.29$   \\  		      
28 & 250.661 & 27.444  & 36.6 & 46 & 38 & 3.36     & $ 35^{+7}_{-7} $	   & $14.42\pm 0.28$   \\  		      
29 & 253.061 & 44.823  & 74.3 & 95 & 41 & 3.10     & $ 83^{+9}_{-10} $     & $14.63\pm 0.28$   \\  		      
30 & 27.115 & 14.038   & 29.3 & 27 & 18 & 5.24     & $ 24^{+4}_{-6} $	   & $14.33\pm 0.30$   \\  		      
31 & 328.925 & 12.525  & 52.1 & 59 & 39 & 5.51     & $ 53^{+6}_{-9} $	   & $14.52\pm 0.28$   \\  		      
32 & 140.964 & 59.512  & 41.5 & 47 & 50 & 7.04     & $ 41^{+7}_{-7} $	   & $14.46\pm 0.29$   \\  		      
33 & 141.833 & 30.232  & 23.5 & 23 & 172 & 28.64   & $ 18^{+4}_{-6} $	   & $14.26\pm 0.30$   \\  		      
34 & 168.213 & 53.856  & 41.3 & 45 & 43 & 8.30     & $ 41^{+6}_{-8} $	   & $14.46\pm 0.28$   \\  		      
35 & 179.130 & 54.361  & 19.3 & 15 & 80 & 38.45    & $ 14^{+3}_{-4} $	   & $14.19\pm 0.29$   \\  		      
36 & 180.201 & -1.188  & 16.4 & 8 & 10 & 8.05	   & $ 8^{+2}_{-4} $	   & $14.03\pm 0.32$   \\  		      
37 & 203.754 & 58.719  & 29.6 & 32 & 27 & 7.13     & $ 29^{+4}_{-7} $	   & $14.37\pm 0.28$   \\  		      
38* & 210.258 & 2.878  & 127.6 & 238 & 115 & 3.28  & $ 204^{+16}_{-16} $   & $14.85\pm 0.31$   & Abell 1835 \\        
39 & 211.664 & 27.600  & 24.9 & 21 & 101 & 19.48   & $ 17^{+4}_{-5} $	   & $14.23\pm 0.29$   \\  		      
40 & 217.705 & 28.153  & 20.8 & 17 & 39 & 11.73    & $ 15^{+4}_{-4} $	   & $14.20\pm 0.29$   \\  		      
42 & 319.704 & 0.560   & 81.7 & 104 & 87 & 5.50    & $ 89^{+10}_{-11} $    & $14.65\pm 0.29$   \\  		      
43 & 354.416 & 0.271   & 89.5 & 139 & 124 & 5.37   & $ 117^{+11}_{-13} $   & $14.73\pm 0.30$   & Abell 2631 (1.4')\\  
\hline          				  
\end{tabular} 					  
\hfill \break
Objects with an ID with an asterisc are not part of the statistical sample, 
because they are X-ray selected. \hfill \break
\end{table*}

%% file: Lx.table.tex
\begin{table}
\caption{X-ray data}
{
\footnotesize
\begin{tabular}{r r r r r r l}
\hline
ID & $t_{exp}$ & obstot & obsbkg & nbox & C & $\log L_X$ \\
     &	&  	&   	& 	&	&  erg s$^{-1}$  \\
\hline
 1   & 125.4 & 259 & 4079 & 33 & 40.41       & $42.52^{0.13}_{-0.06}$  \\    
 2   & 4.3   & 35 & 269 & 22 & 41.49	     & $42.84^{0.14}_{-0.10}$  \\    
 3   & 8.9   & 457 & 356 & 51 & 41.73	     & $44.38^{0.02}_{-0.02}$	\\   
 4*  & 36.4  & 1223 & 1534 & 57 & 41.31      & $44.38^{0.01}_{-0.01}$  \\    
 6   & 5.0   & 156 & 325 & 52 & 41.77	     & $43.95^{0.04}_{-0.04}$	\\   
 7   & 4.2   & 53 & 256 & 25 & 41.38	     & $43.02^{0.08}_{-0.07}$  \\    
 9   & 184.1 & 415 & 3827 & 10 & 39.67       & $41.39^{0.48}_{-0.14}$  \\    
10  & 3.8   & 59 & 190 & 41 & 41.80	     & $43.54^{0.06}_{-0.06}$  \\    
11  & 94.1  & 231 & 4174 & 35 & 40.26	     & $42.29^{0.15}_{-0.06}$  \\    
12  & 121.1 & 176 & 4509 & 63 & 40.63	     & $42.64^{0.10}_{-0.06}$  \\    
13  & 149.3 & 150 & 4503 & 61 & 40.66	     & $42.53^{0.14}_{-0.07}$  \\    
16  & 10.2  & 204 & 330 & 13 & 41.04	     & $43.30^{0.04}_{-0.03}$  \\    
17  & 5.0   & 296 & 171 & 29 & 41.68	     & $44.14^{0.03}_{-0.02}$  \\    
19  & 67.7  & 277 & 2326 & 34 & 40.59	     & $42.91^{0.05}_{-0.04}$  \\    
20  & 4.4   & 567 & 188 & 10 & 41.30	     & $44.04^{0.02}_{-0.02}$  \\    
21*  & 30.1  & 3522 & 2133 & 62 & 41.38      & $44.93^{0.01}_{-0.01}$  \\    
23  & 4.2   & 101 & 562 & 58 & 41.88	     & $43.85^{0.05}_{-0.05}$  \\    
24  & 2.1   & 179 & 239 & 27 & 41.84	     & $44.07^{0.03}_{-0.03}$  \\    
25  & 5.0   & 59 & 400 & 43 & 41.74	     & $43.44^{0.07}_{-0.07}$  \\    
26  & 12.4  & 223 & 337 & 19 & 41.20	     & $43.51^{0.03}_{-0.03}$  \\    
27  & 4.5   & 190 & 340 & 65 & 41.97	     & $44.24^{0.03}_{-0.03}$  \\    
28  & 4.6   & 310 & 190 & 14 & 41.43	     & $43.90^{0.03}_{-0.03}$  \\    
29  & 3.6   & 101 & 262 & 48 & 41.90	     & $43.88^{0.05}_{-0.04}$  \\    
30  & 17.4  & 434 & 840 & 17 & 40.94	     & $43.52^{0.03}_{-0.02}$  \\    
31  & 9.9   & 691 & 396 & 36 & 41.60	     & $44.43^{0.02}_{-0.02}$  \\    
32  & 18.0  & 49 & 724 & 52 & 41.29	     & $42.83^{0.10}_{-0.08}$  \\    
33  & 159.3 & 380 & 4650 & 61 & 40.80	     & $43.28^{0.04}_{-0.03}$  \\    
34  & 7.8   & 38 & 468 & 77 & 41.72	     & $43.23^{0.09}_{-0.08}$  \\    
35  & 89.9  & 144 & 2539 & 58 & 40.76	     & $42.76^{0.08}_{-0.05}$  \\    
36  & 26.8  & 79 & 828 & 15 & 40.59	     & $41.92^{0.33}_{-0.11}$  \\    
37  & 4.8   & 53 & 199 & 25 & 41.63	     & $43.28^{0.08}_{-0.06}$  \\    
38*  & 12.9  & 3135 & 1257 & 76 & 41.80      & $45.29^{0.01}_{-0.01}$  \\    
39  & 81.2  & 412 & 3302 & 66 & 40.87	     & $43.43^{0.03}_{-0.03}$  \\    
40  & 27.3  & 50 & 1098 & 28 & 40.91	     & $41.94^{0.43}_{-0.13}$  \\    
42  & 6.3   & 33 & 387 & 98 & 42.13	     & $43.60^{0.09}_{-0.08}$  \\    
43  & 2.9   & 96 & 198 & 96 & 42.51	     & $44.48^{0.05}_{-0.04}$  \\    
\hline \end{tabular}
\hfill \break
Objects with an ID with an asterisc are not part of the statistical sample, 
because they are X-ray selected. \hfill \break
}  
\end{table}